\documentclass[epj]{svjour}
\usepackage{graphics}

\usepackage{graphicx}% Include figure files
\usepackage{dcolumn}% Align table columns on decimal point
\usepackage{bm}% bold math
\usepackage{amssymb}
\usepackage{amsfonts}
\usepackage{amsmath}
\usepackage{latexsym}
\usepackage{dsfont}
\usepackage{multirow}
\usepackage{color}
\usepackage[mathscr]{eucal}
\usepackage{subfigure}
\usepackage{bm}
\usepackage{slashed}
\usepackage{multicol}
\usepackage{blindtext}

\newcommand{\ud}{\mathrm{d}}

\newcommand{\uTr}{\mathrm{Tr}}

\newcommand{\uslash}{/\!\!\!\!}

\newcommand{\uvec}[1]{\vec{#1}}
\newcommand{\pure}{\text{pure}}
\newcommand{\phys}{\text{phys}}

\def\rmsmall #1{\mbox{\scriptsize #1}}

\newcommand{\bea}{\begin{eqnarray}}
\newcommand{\eea}{\end{eqnarray}}
\newcommand{\be}{\begin{equation}}
\newcommand{\ee}{\end{equation}}

\newcommand{\ba}{\begin{eqnarray}}
\newcommand{\ea}{\end{eqnarray}}

\definecolor{green}{rgb}{0,.5,0}

\usepackage[normalem]{ulem}

\renewcommand\sout{\bgroup \color[rgb]{0.55,0.00,0.99} \ULdepth=-.5ex \ULset}

\begin{document}

\title{The Parton Orbital Angular Momentum: Status and Prospects}

\author{Keh-Fei Liu\inst{1} \and C\'edric Lorc\'e\inst{2,3}% etc
}                     % Do not remove
\institute{Department of Physics and Astronomy, University of Kentucky, Lexington, KY 40506 USA \and SLAC National Accelerator Laboratory, Stanford 
  University, Menlo Park, CA 94025 USA \and IFPA,  AGO Department, Universit\'e de Li\` ege, Sart-Tilman, 4000 Li\`ege, Belgium }
\abstract{Theoretical progress on the formulation and classification of the quark and gluon orbital angular momenta (OAM) is reviewed.  Their relation to parton distributions and open questions and puzzles are discussed. 
We give a status report on the lattice calculation of the parton kinetic and canonical OAM and point out several strategies to calculate 
the quark and gluon canonical OAM on the lattice.
\PACS{{12.38.-t}{Quantum chromodynamics}\and
      {14.20.Dh}{Protons and neutrons}   \and
      {12.38.Gc}{Lattice QCD calculations}
     } % end of PACS codes
} %end of abstract
\maketitle
\section{Introduction}
\label{sec1}

One of the key questions in hadron physics is to understand how the spin of the proton originates from the parton spin and orbital motion. Based on the naive quark model picture, supported among others by its success in hadron spectroscopy, it was expected that the quark spin contribution should account for most\footnote{In relativistic quark models, the quark orbital motion can contribute significantly to the proton spin budget, but the quark spin remains by far the dominant contribution~\cite{Burkardt:2010he}.} of the proton spin. Deep-inelastic scattering experiments have been conducted and obtained the surprising result that the quark spin contribution is actually quite small, about $25\%$~\cite{Aidala:2012mv}. Relativistic quark models argue that the quark orbital motion reduces significantly the quark spin contribution, but in practice this effect falls short in explaining such a small value. According to Quantum ChromoDynamics (QCD), gluon polarization also contributes to the proton spin budget. This contribution appears to be of the order of $35\%$~\cite{deFlorian:2014yva,Nocera:2014gqa}, leaving about $40\%$ which should come from the orbital angular momentum (OAM) of quarks and gluons. This number is large and reflects the relativistic nature of the quark-gluon bound state. It also stresses the paramount importance of actually measuring the OAM contribution.

The parton OAM is more difficult to access than the parton helicity owing to the fact that it measures a correlation between position and momentum, and therefore requires some information about parton phase space. Moreover, it also depends on how the total AM is divided into separate quark and gluon contributions which is intrinsically ambiguous due to quark-gluon couplings. It has long been thought that only the so-called kinetic decomposition of the proton spin makes sense because its quark and gluon contributions can be extracted from experiments and computed on the lattice. Recent theoretical and experimental progress have however shown that the canonical decomposition can also be accessed experimentally and computed on the lattice, though in a more complicated way. Kinetic and canonical decomposition are two complementary decompositions of the nucleon spin with their own advantages and disadvantages. For more detailed discussions about the angular momentum (AM) decomposition in gauge theories, see the recent reviews~\cite{Leader:2012md,Wakamatsu:2014zza}. For a theoretical and experimental overview of optical AM, see~\cite{Bliokh:2015doa}.

The aim of the present paper is to summarize critically the current theoretical knowledge about the parton OAM and give some prospects for future interesting developments. The paper is organized as follow. We quickly review in sec.~\ref{sec2} the recent theoretical development about the proton spin decomposition and discuss both kinetic and canonical decompositions. In Sec.~\ref{sec3}, we show how the different contributions, and in particular the parton OAM, can be expressed in terms of parton distributions and therefore how they can be accessed experimentally. We then discuss in detail in Sec.~\ref{sec4} several issues and puzzles related to the definition and calculation of OAM. We also comment critically various claims found in the literature. In Sec.~\ref{sec5}, we present in a nutshell some lattice calculations of the various AM contributions, starting first with the Mellin moment approach followed then by the more recent large momentum approach. We suggest in Sec.~\ref{sec6} that a realistic effective theory of baryons consistent with chiral symmetry could provide valuable quantitative insights concerning the question of the proton spin decomposition. In Sec.~\ref{sec7} we present the concept of spin-orbit correlation and show that it is closely related to parton OAM. In particular, all the discussions and developments regarding OAM find a natural counterpart when studying the spin-orbit correlation. Finally, we summarize in Sec.~\ref{sec8}.

\section{The gauge-invariant angular momentum decomposition}\label{sec2}

Generically, the total AM operator in QCD can be decomposed into five gauge-invariant contributions~\cite{Wakamatsu:2010qj}
\begin{equation}\label{decomposition}
\uvec J=\uvec S^q+\uvec S^G+\uvec L^q_\text{kin}+\uvec L^G_\text{can}+\uvec L_\text{pot},
\end{equation}
where
\begin{equation}\label{dec5}
\begin{aligned}
\uvec S^q&=\int\ud^3r\,\psi^\dag\tfrac{1}{2}\uvec\Sigma\psi,\\
\uvec S^G&=\int\ud^3r\,\uvec E^a\times\uvec A^a_\phys,\\
\uvec L^q_\text{kin}&=\int\ud^3r\,\psi^\dag(\uvec r\times i\uvec D)\psi,\\
\uvec L^G_\text{can}&=-\int\ud^3r\,E^{aj}(\uvec r\times\uvec{\mathcal D}^{ab}_\pure) A^{bj}_\phys,\\
\uvec L_\text{pot}&=-\int\ud^3r\,\rho^a\,\uvec r\times\uvec A^a_\phys
\end{aligned}
\end{equation}
are the quark spin, gluon spin, quark kinetic\footnote{The kinetic OAM is sometimes called mechanical or dynamical OAM in the literature.} OAM, gluon canonical OAM and potential OAM, respectively. 

Following Chen \emph{et al.}~\cite{Chen:2008ag,Chen:2009mr}, the gauge potential has been decomposed into a pure-gauge field and a ``physical'' field
\begin{equation}\label{Adec}
A^\mu(r)=A^\mu_\pure(r)+A^\mu_\phys(r)
\end{equation}
such that
\begin{equation}\label{puregaugecond}
F^{\mu\nu}_\pure=\partial^\mu A^\nu_\pure-\partial^\nu A^\mu_\pure-ig[A^\mu_\pure,A^\nu_\pure]=0.
\end{equation}
Under a gauge transformation, $A^\mu_\pure$ is assumed to transform as a connection, \emph{i.e.} like the gauge potential $A$
\begin{equation}   \label{inhomo}  
A^\mu_\pure(r)\mapsto g(r)[A^\mu_\pure(r)+\tfrac{i}{g}\partial^\mu]g^{-1}(r).
\end{equation}
It follows that $A^\mu_\phys$, which is the difference between two connections, transforms homogeneously
\begin{equation}   \label{homo}
A^\mu_\phys(r)\mapsto g(r)A^\mu_\phys(r)g^{-1}(r).
\end{equation}
The pure-gauge covariant derivatives are defined like the ordinary covariant derivatives with the gauge potential $A^\mu$ replaced by the pure-gauge field $A^\mu_\pure$
\begin{equation}
\begin{aligned}
D^\mu_\pure&=\partial^\mu-igA^\mu_\pure,\\
\mathcal D^\mu_\pure&=\partial^\mu-ig[A^\mu_\pure,\quad].
\end{aligned}
\end{equation}
This ensures the explicit gauge invariance of the expressions in Eq.~\eqref{dec5}. It turns out that the Chen \emph{et al.} approach is just a particular case of the standard background field method with the background field $A_\pure$ chosen to be flat~\cite{Lorce:2013gxa,Lorce:2013bja}, \emph{i.e.} satisfying the pure-gauge condition in Eq.~\eqref{puregaugecond}.

Because of the QCD equations of motion
\begin{equation}
\rho^a=g\psi^\dag t^a\psi=\uvec{\mathcal D}^{ab}\cdot\uvec E^b,
\end{equation}
the interpretation of the potential OAM as a quark or gluon contribution is ambiguous. It represents in some sense the change in OAM as the quark moves through the color field created by the spectator system~\cite{Burkardt:2012sd}. When the potential OAM is combined with the quark kinetic OAM, one obtains the quark canonical OAM
\begin{equation}\label{potOAM}
\begin{aligned}
\uvec L^q_\text{can}&=\uvec L^q_\text{kin}+\uvec L_\text{pot}\\
&=\int\ud^3r\,\psi^\dag(\uvec r\times i\uvec D_\pure)\psi,
\end{aligned}
\end{equation}
leading to the so-called gauge-invariant canonical or Chen \emph{et al.} decomposition~\cite{Chen:2008ag,Chen:2009mr}
\begin{equation}   \label{chen_decomp}
\uvec J=\uvec S^q+\uvec S^G+\uvec L^q_\text{can}+\uvec L^G_\text{can}
\end{equation}
which can be seen as a gauge-invariant version\footnote{By gauge-invariant extension, we mean a gauge-invariant quantity which gives the same physical results as a gauge-variant one considered in a certain fixed gauge.} (or extension) of the Jaffe-Manohar decomposition~\cite{Jaffe:1989jz}. 
When the potential OAM is combined with the gluon canonical OAM, one obtains the gluon kinetic OAM
\begin{equation}\label{LGkin}
\begin{aligned}
\uvec L^G_\text{kin}&=\uvec L^G_\text{can}+\uvec L_\text{pot}\\
&=\int\ud^3r\,\uvec r\times[(\uvec A^a_\phys\times\uvec{\mathfrak D}^{ab}_\pure)\times \uvec E^b]
\end{aligned}
\end{equation}
with $\mathfrak D^\mu_\pure=\tfrac{1}{2}(\mathcal D^\mu+\mathcal D^\mu_\pure)$~\cite{Lorce:2013fpa},
leading to the so-called gauge-invariant kinetic or Wakamatsu decomposition~\cite{Wakamatsu:2010qj,Wakamatsu:2010cb}
\begin{equation}
\uvec J=\uvec S^q+\uvec S^G+\uvec L^q_\text{kin}+\uvec L^G_\text{kin}
\end{equation}
which generalizes the Ji decomposition~\cite{Ji:1996ek}
\begin{equation}  \label{ji_decomp}
\uvec J=\uvec S^q+\uvec L^q_\text{kin}+\uvec J^G_\text{kin},
\end{equation}
where  
%### Comment: I have promoted this to an equation.
\begin{equation}  \label{JG}
\uvec J^G_\text{kin}=\uvec S^G+\uvec L^G_\text{kin}=\int\ud^3r\,[\uvec r\times(\uvec E^a\times\uvec B^a)]
\end{equation}
involves $\uvec E^a\times\uvec B^a$, the Poynting vector of the glue.
\newline

The gauge invariance of the terms in Eq.~\eqref{dec5} seems to contradict the textbook claim that a gauge-invariant decomposition of the gluon total AM into spin and OAM contributions is impossible~\cite{Berestetskii,Jauch}. The contradiction is only apparent though since textbooks deal actually with local fields, whereas $A^\mu_\pure$ and $A^\mu_\phys$ are intrinsically non-local\footnote{The fields $A^\mu_\pure$ and $A^\mu_\phys$ can be expressed solely in terms of the full gauge potential $A^\mu$, but this requires the knowledge of $A^\mu$ at least over a finite neighborhood.}~\cite{Lorce:2012rr,Lorce:2012ce}. As a result, manifest Lorentz covariance is lost. Moreover, the pure-gauge condition~\eqref{puregaugecond} is not sufficient for fixing uniquely the form of $A^\mu_\pure$ and $A^\mu_\phys$, leading to a new freedom at the level of the QCD Lagrangian
\begin{equation}\label{Stueckelberg}
\begin{aligned}
A^\mu_\pure(r)&\mapsto A^\mu_\pure(r)+\partial^\mu C(r),\\
A^\mu_\phys(r)&\mapsto A^\mu_\phys(r)-\partial ^\mu C(r),
\end{aligned}
\end{equation}
referred to as the Stueckelberg transformation~\cite{Stoilov:2010pv,Lorce:2012rr,Francois:2014bya} and related to the concept of background dependence \cite{Lorce:2013bja}. In theory, this freedom is not welcome as it makes the initial decomposition~\eqref{decomposition} intrinsically ambiguous. In practice, however, the experimental conditions usually provide us with a natural choice for the background field $A^\mu_\pure$~\cite{Bashinsky:1998if,Lorce:2012rr,Wakamatsu:2014toa}, giving \emph{e.g.} a preferred spatial direction and explaining why the photon spin and OAM are routinely measured in optics\footnote{The direction of propagation of a laser beam provides a natural direction along which one can consistently define and measure separately the photon spin and OAM projections.}~\cite{VanEnk:1994,Bliokh:2014ara,Bliokh:2015doa}. In high-energy experiments where the proton internal structure is probed, the off-shell probe provides the natural direction~\cite{Collins:2011zzd}. In other words, even if the fundamental Lagrangian is Stueckelberg invariant (\emph{i.e.} background independent), physical quantities extracted within a controlled expansion scheme are often associated with a natural fixed background.
\newline

In summary, a complete gauge-invariant decomposition of the proton spin makes sense and is feasible, but its precise form depends on how one probes the system. There is also an intrinsic ambiguity in decomposing the total OAM into quark and gluon contributions, leading to essentially two types of decompositions: canonical and kinetic. This ambiguity can however be avoided by refraining from interpreting the potential OAM as either quark or gluon contribution.

\section{Relations to parton distributions}\label{sec3}

Using the Belinfante-Rosenfeld version of the QCD energy-momentum tensor, Ji~\cite{Ji:1996ek} derived a remarkable relation between the quark/gluon total kinetic AM\footnote{This relation holds for a spin-$1/2$ target. For a spin-$1$ target, a similar relation has been obtained by Taneja \emph{et al.}~\cite{Taneja:2011sy}.} and twist-2 generalized parton distributions (GPDs)
\begin{equation}\label{Jirel}
\langle J^{q,G}_\text{kin}\rangle=\tfrac{1}{2}\int\ud x\,x[H^{q,G}(x,0,0)+E^{q,G}(x,0,0)].
\end{equation}
This relation has been obtained for the longitudinal component of the total AM $J_L=\uvec J\cdot\uvec P/|\uvec P|$ and does not depend on the magnitude of the proton momentum $|\uvec P|$~\cite{Ji:1997pf,Leader:2012md}. The corresponding relation for the transverse component of the total AM $J_T$ has been obtained by Leader~\cite{Leader:2011cr} and is momentum dependent, a consequence of the fact that longitudinal boosts and transverse rotations do not commute. Only in the rest frame $|\uvec P|\to 0$ do the relations for the longitudinal and transverse components coincide, owing to rotational symmetry. Interestingly, Burkardt~\cite{Burkardt:2005hp} showed that the Ji relation~\eqref{Jirel} has a simple interpretation when the transverse component of the total AM is considered in the rest frame. In order to obtain a momentum-independent relation for the transverse polarization, Ji, Xiong and Yuan~\cite{Ji:2012vj} proposed to use the transverse component of the Pauli-Lubanski vector instead of the total AM, but even in this case the separate quark and gluon contributions remain frame dependent~\cite{Leader:2012ar,Leader:2013jra,Hatta:2012jm,Harindranath:2013goa}. From now on, we will implicitly choose the $z$ axis along the proton momentum and restrict ourselves to the longitudinal component of AM.

The quark/gluon kinetic OAM is then obtained by subtracting the quark/gluon spin contribution, which is given in 
the $\overline{MS}$ scheme by the first Mellin moment of the quark/gluon helicity distribution
\begin{equation}\label{helicityrel}
\begin{aligned}
\langle S^q\rangle&=\tfrac{1}{2}\int\ud x\,\Delta q(x),\\
\langle S^G\rangle&=\int\ud x\,\Delta G(x),
\end{aligned}
\end{equation}
from the quark/gluon total kinetic AM
\begin{equation}\label{JiOAMrel}
\langle L^{q,G}_\text{kin}\rangle=\langle J^{q,G}_\text{kin}\rangle-\langle S^{q,G}\rangle.
\end{equation}
Penttinen, Polyakov, Shuvaev and Strikman (PPSS)~\cite{Penttinen:2000dg} showed that the quark kinetic OAM can also be directly expressed in terms of a two-parton twist-3 GPD~\cite{Kiptily:2002nx,Hagler:2003jw,Hatta:2012cs,Lorce:2015lna}
\begin{equation}\label{twist3rel}
\langle L^q_\text{kin}\rangle=-\int\ud x\,x\,G^q_2(x,0,0)
\end{equation}
which enters the description of the longitudinal target spin asymmetry of deeply virtual Compton scattering~\cite{Courtoy:2013oaa}. The quark canonical OAM can also in principle be accessed \emph{via} twist-3 GPDs by extracting the potential OAM $\langle L_\text{pot}\rangle$ and adding it to the kinetic OAM $\langle L^q_\text{kin}\rangle$, see Eq.~\eqref{potOAM}. Explicit expressions can be found in Refs.~\cite{Hatta:2012cs,Ji:2012ba}.
\newline

Since GPDs were shown to give access to AM and many transverse-momentum distributions (TMDs) would be identically zero in absence of parton OAM, there is some hope that TMDs could also give quantitative information about OAM. Based on quark model calculations, the following relations have been proposed~\cite{Ma:1998ar,She:2009jq,Avakian:2010br,Efremov:2010cy}
\begin{equation}\label{LzTMD}
\begin{aligned}
\langle L_\text{can}^q\rangle
&=\int\ud x\,\ud^2k_\perp\left[h^q_1(x,\uvec k^2_\perp)-g^q_{1L}(x,\uvec k^2_\perp)\right],\\
&=-\int\ud x\,\ud^2k_\perp\,\tfrac{\uvec k_\perp^2}{2M^2}\,h_{1T}^{\perp q}(x,\uvec k^2_\perp).
\end{aligned}
\end{equation}
Unfortunately, these relations cannot be derived directly from QCD and turn out to be valid only in a restricted class of models~\cite{Lorce:2011dv,Lorce:2011kn,Lorce:2011zta}. In general, no direct quantitave relations between OAM and TMDs should actually be expected as the former represents a correlation between parton position and momentum whereas the latter only provide information about the momentum distribution. Nevertheless, TMDs do provide some indirect information about the OAM content of the nucleon by constraining the wave function of the latter.

Another possibility 
%relating indirectly OAM to TMDs 
has also been explored. Burkardt~\cite{Burkardt:2002ks,Burkardt:2003uw} suggested that the quark Sivers TMD $f_{1T}^{\perp q}(x,\uvec k^2_\perp)$ and the quark GPD $E^q(x,\xi,t)$ could be related by
a chromodynamic lensing mechanism 
\begin{equation}
\begin{aligned}
&\int\ud^2k_\perp\,\tfrac{\uvec k^2_\perp}{2M^2}\,f_{1T}^{\perp q}(x,\uvec k^2_\perp)\\ &\qquad\qquad\propto
\int \ud^2b_\perp \,\bar{\mathcal I}(\uvec b_\perp)
\,(\uvec S_T\times\uvec \partial_{b_\perp})_z\, {\cal E}^q(x, \uvec b_\perp^2),
\end{aligned}
\end{equation}
where $\bar{\mathcal I}(\boldsymbol b_\perp)$ is called the lensing function and ${\cal E}^q(x,\boldsymbol{b}_\perp^2) =\int \frac{d^2 \Delta_\perp}{(2\pi)^2} \, e^{-i\boldsymbol{b}_\perp \cdot \boldsymbol{\Delta}_\perp}\, E^q(x,0,-\boldsymbol{\Delta}_\perp^2)$. This relation is supported by some model calculations~\cite{Burkardt:2003uw,Burkardt:2003je,Meissner:2007rx,Gamberg:2009uk} but can hardly be put on firmer theoretical grounds. A variation of it has however been used by Bacchetta and Radici~\cite{Bacchetta:2011gx} to fit SIDIS data for the Sivers effect with the integral constrained by the anomalous magnetic moments which led to a new estimate of $\langle J^{q}_\text{kin}\rangle$ in good agreement with most common GPD extractions~\cite{Guidal:2004nd,Diehl:2004cx,Ahmad:2006gn,Goloskokov:2008ib,Diehl:2013xca}.
\newline

As shown by Lorc\'e and Pasquini~\cite{Lorce:2011kd,Lorce:2011ni}, the most intuitive expression for the quark/gluon OAM is as a phase-space integral \begin{equation}\label{OAMWigner}
\begin{aligned}
&\langle L^{q,G}_{\mathcal W}\rangle=\\
&\quad\int\ud x\,\ud^2k_\perp\,\ud^2b_\perp\,(\vec b_\perp\times\vec k_\perp)_z\,\rho^{q,G}_{++}(x,\vec k_\perp,\vec b_\perp;\mathcal W),
\end{aligned}
\end{equation}
where the Wilson line $\mathcal W$ ensures gauge invariance and the relativistic phase-space (or Wigner) distribution $\rho^{q,G}_{++}$ can be interpreted semi-classically as giving the quasi-probability\footnote{The relativistic phase-space distribution is generally not positive-definite because of the Heisenberg's uncertainty principle. At best it can be seen as a weighing function in phase-space. Alternatively, one can consider the positive-definite Husimi distribution but one loses the probabilistic interpretation of the corresponding position and momentum-space distributions~\cite{Hagiwara:2014iya}.} for finding at the transverse position $\uvec b_\perp$ a quark/gluon with momentum $(xP^+,\uvec k_\perp)$ inside a longitudinally polarized ($\Lambda'=\Lambda=+$) proton. Unlike the version proposed in Refs.~\cite{Ji:2003ak,Belitsky:2003nz}, this phase-space representation does not require any relativistic correction because it is based on the Euclidean subgroup of the light-front formalism~\cite{Soper:1976jc,Burkardt:2000za,Burkardt:2005hp}. 

Under a Fourier transform, the phase-space distribution is related to the generalized transverse-momentum dependent distributions (GTMDs)~\cite{Meissner:2009ww,Lorce:2011dv,Lorce:2013pza}, leading to the simple relation~\cite{Lorce:2011kd,Hatta:2011ku,Kanazawa:2014nha}
\begin{equation}\label{LzGTMD}
\langle L^{q,G}_{\mathcal W}\rangle=-\int\ud x\,\ud^2k_\perp\,\tfrac{\uvec k^2_\perp}{M^2}\,F^{q,G}_{14}(x,0,\uvec k_\perp,\uvec 0_\perp;\mathcal W).
\end{equation}
The shape of the Wilson line determines the type of OAM \cite{Ji:2012sj,Lorce:2012ce,Burkardt:2012sd}. For a staple-like Wilson line $\mathcal W_{\sqsupset}$, like \emph{e.g.} the one involved in the description of semi-inclusive deep-inelastic scattering and Drell-Yan process~\cite{Collins:2011zzd}, Eq.~\eqref{LzGTMD} leads to the canonical version of the quark/gluon OAM $\langle L^{q,G}_\text{can}\rangle=\langle L^{q,G}_{\mathcal W_{\sqsupset}}\rangle$ irrespective of whether the staple is future or past-pointing~\cite{Hatta:2011ku,Hatta:2012cs,Ji:2012ba,Lorce:2012ce,Lorce:2015lna}. For a straight Wilson line $\mathcal W_\mid$, it leads to the kinetic version of the quark OAM $\langle L^q_\text{kin}\rangle=\langle L^q_{\mathcal W_\mid}\rangle$~\cite{Ji:2012sj} and the Ji-Xiong-Yuan (JXY)~\cite{Ji:2012ba} definition of the gauge-invariant gluon OAM $\langle L^G_\text{JXY}\rangle=\langle L^G_{\mathcal W_\mid}\rangle$. Note that while the gluon kinetic OAM is given by Eq.~\eqref{LGkin}, the gluon JXY OAM is defined as  
\begin{equation}
\begin{aligned}
\uvec L^G_\text{JXY}&=-\int\ud^3r\,E^{aj}(\uvec r\times\uvec{\mathcal D}^{ab}) A^{bj}_\phys\\
&=\uvec L^G_\text{can}+\uvec L^G_\text{pot},
\end{aligned}
\end{equation}
where
\begin{equation}
\begin{aligned}
\uvec L^G_\text{pot}&=gf^{abc}\int\ud^3r\,E^{aj}A^{bj}_\phys\,(\uvec r\times\uvec A^{c}_\phys) \\
&\neq \uvec L_\text{pot}.
\end{aligned}
\end{equation}
The JXY decomposition of the proton spin then reads
\begin{equation}\label{JXYdecomp}
\uvec J=\uvec S^q+\uvec S^G+\uvec L^q_\text{JXY}+\uvec L^G_\text{JXY}-\uvec L^q_\text{pot}-\uvec L^G_\text{pot}
\end{equation}
with $\uvec L^q_\text{JXY}=\uvec L^q_\text{kin}$ and $\uvec L^q_\text{pot}=-\uvec L_\text{pot}$.

While GTMDs provided for the first time a clear relation between canonical OAM and parton distributions, it is not known so far how to extract them from actual experiments, except possibly at small $x$~\cite{Meissner:2009ww}. Moreover, the GTMD $F_{14}$ does not reduce to any GPD or TMD, and so cannot be directly constrained. However, recent theoretical developments~\cite{Ji:2013fga,Ji:2014lra} and encouraging results~\cite{Sufian:2014jma} in the framework of large-momentum effective field theory (LaMET)~\cite{Ji:2013dva} open the interesting possibility of computing GTMDs and OAM directly on the lattice, see section~\ref{LaMETsec}. Moreover, effective models, constrained by experimental data, also provide indirect access to GTMDs and OAM~\cite{Lorce:2011kd,Lorce:2011ni,Kanazawa:2014nha,Mukherjee:2014nya,Mukherjee:2015aja,Liu:2015eqa}.

\section{Current theoretical questions and puzzles}\label{sec4}

From a theoretical point of view, it is naturally very interesting to 
check within models the different relations presented in the previous section together with the AM sum rule. Several results have already been obtained in relativistic quark models and perturbative models, but the situation is rather confusing and definitely requires further investigations.

In Ref.~\cite{Lorce:2011kd}, Lorc\'e and Pasquini compared, within two different relativistic light-front quark models, the value for the kinetic OAM obtained from Eq.~\eqref{JiOAMrel} and the Ji relation~\eqref{Jirel} with the value for the canonical OAM obtained from the model-dependent TMD relations~\eqref{LzTMD} and the model-independent GTMD relation~\eqref{LzGTMD}, see Table~\ref{OAMtable}. All these three expressions agree on the total OAM, showing that the conservation of total AM is correctly implemented in the models. They however differ in the split of the latter into the various flavor contributions. For the canonical OAM obtained from Eqs.~\eqref{LzTMD} and~\eqref{LzGTMD}, this can be attributed to the difference in the definition of the ``center'' of the target about which the OAM is computed~\cite{Lorce:2011kn}. The difference in the flavor split between canonical and kinetic OAM is particularly puzzling because it is usually believed that the two forms of OAM should coincide in quark models because of the absence of explicit gluon degrees of freedom. Interestingly, a similar observation has been made earlier by Wakamatsu and Tsujimoto~\cite{Wakamatsu:2005vk} within the chiral quark-soliton model based on the large-$N_c$ picture of QCD, and recently by Liu and Ma~\cite{Liu:2014zla} within the axial-vector diquark model. So far no satisfactory explanation of these results has been proposed.

\begin{table}[t]
\begin{center}
\caption{\footnotesize{Kinetic and canonical quark OAM obtained in the light-front constituent quark model (LFCQM) and the light-front chiral-quark soliton model (LF$\chi$QSM)~\cite{Lorce:2011kd}.}}\label{OAMtable}
\begin{tabular}{c|ccc}\hline\noalign{\smallskip}
LFCQM&$u$&$d$&Total\\
\noalign{\smallskip}\hline\noalign{\smallskip}
$\langle L^q_\text{kin}\rangle_\text{Eq.~\eqref{JiOAMrel}}$&$0.071$&$0.055$&$0.126$\\
$\langle L^q_\text{can}\rangle_\text{Eq.~\eqref{LzTMD}}$&$0.131$&$-0.005$&$0.126$\\
$\langle L^q_\text{can}\rangle_\text{Eq.~\eqref{LzGTMD}}$&$0.169$&$-0.042$&$0.126$\\
\noalign{\smallskip}\hline\noalign{\smallskip}
LF$\chi$QSM&$u$&$d$&Total\\
\noalign{\smallskip}\hline\noalign{\smallskip}
$\langle L^q_\text{kin}\rangle_\text{Eq.~\eqref{JiOAMrel}}$&$-0.008$&$0.077$&$0.069$\\
$\langle L^q_\text{can}\rangle_\text{Eq.~\eqref{LzTMD}}$&$0.073$&$-0.004$&$0.069$\\
$\langle L^q_\text{can}\rangle_\text{Eq.~\eqref{LzGTMD}}$&$0.093$&$-0.023$&$0.069$\\
\noalign{\smallskip}\hline
\end{tabular}
\end{center}
\end{table}

In the scalar diquark model, the situation is also quite puzzling since, on the one hand, Burkardt and BC~\cite{Burkardt:2008ua} found that the quark canonical and kinetic OAM do coincide using the Pauli-Villars regularization while, on the other hand, Liu and Ma~\cite{Liu:2014zla} found that they differ using a form factor regularization. For an electron in QED to order $\alpha$, the difference between electron canonical and kinetic OAM is non-zero, as expected for a gauge theory. It turns out to be finite but the precise value depends on the regularization prescription~\cite{Burkardt:2008ua,Liu:2014fxa}. Since this difference is attributed to the potential OAM~\eqref{potOAM}, Liu and Ma~\cite{Liu:2014fxa} computed explicitly the latter and found the suprising result that it does not account for the observed difference. Lowdon~\cite{Lowdon:2014dda} emphasized that surface terms usually assumed to vanish in field theory could actually contribute. Liu and Ma then computed explicitly the contribution from a surface term and found it to be non-zero\footnote{This is actually expected since plane waves are somewhat pathological as they do not vanish at infinity, see~\cite{Leader:2012md}.}. Unfortunately, the combination of potential OAM and surface term is still not able to explain the difference between canonical and kinetic OAM, irrespective of the adopted regularization prescription.

Obviously, further investigations are needed to clarify the situation. In particular, explicit calculations of the PPSS relation~\eqref{twist3rel} and the GTMD relation~\eqref{LzGTMD} with different Wilson lines should shed some light. Also, special attention to Lorentz covariance, equations of motion, regularization prescription and surface terms should help resolve the current puzzles.
\newline

So far, we have focused our discussions on the AM integrated over all phase space. It is however possible to define densities of AM, and therefore map the latter in position and momentum space. There is however a lot of confusion due to the large number of definitions proposed in the literature. One of the reasons for this confusion is that densities differing by a so-called superpotential term $\partial_\alpha X^{[\alpha\mu]\cdots}(r)$, where square brackets mean antisymmetrization of indices, lead to the same integrated quantities. The archetypical example is the total AM of a system, which can be obtained either from the canonical density or the Belinfante-improved density. In the former case, the AM density is naturally decomposed into an OAM density and a spin density, while in the latter case, the spin density is converted into an additional contribution to OAM density plus a superpotential term. One has to be careful when defining AM at the density level as superpotentials modify the distribution and in turn destroy the interpretation as AM density. It is therefore crucial to keep track of these superpotential terms when discussing AM at the density level.

A first approach has been proposed by H\"agler and Sch\"afer~\cite{Hagler:1998kg} where the higher moments of the canonical AM densities in the light-front gauge $A^+=0$ are defined by a tower of generalized canonical AM operators, like \emph{e.g.} 
\begin{equation}
L^{q,\mu_1\cdots\mu_n}_\text{can}=\int\ud^3r\,\overline\psi\gamma^+(\uvec r_\perp\times i\uvec\partial_\perp)_z\,iD^{\mu_1}\cdots iD^{\mu_n}\psi.
\end{equation}
Hoodbhoy, Ji and Lu~\cite{Hoodbhoy:1998yb} adopted the same strategy to define the corresponding higher moments of the kinetic AM densities and concluded that
\begin{equation}\label{kindens}
\langle L^q_\text{kin}\rangle(x)=\tfrac{x}{2}\left[H^q(x,0,0)+E^q(x,0,0)\right]-\tfrac{1}{2}\Delta q(x).
\end{equation}
Note however that $\langle L^q_\text{kin}\rangle(x)$ is not defined from the simple tower
\begin{equation}\label{qkinOAMmom}
\begin{aligned}
&\tilde L^{q,+\cdots +}_\text{kin}=\\
&\tfrac{1}{n+1}\sum_{j=0}^{n}\int\ud^3r\,\overline\psi\gamma^+(iD^+)^j(\uvec r_\perp\times i\uvec D_\perp)_z\,(iD^+)^{n-j}\psi
\end{aligned}
\end{equation}
but involves another complicated tower $\Delta L^{+\cdots +}$ such that the sum $L^{q,+\cdots +}_\text{kin}=\tilde L^{q,+\cdots +}_\text{kin}+\Delta L^{+\cdots +}$ evolves as a leading-twist operator. Moreover, as noted in Ref.~\cite{Leader:2013jra}, Hoodbhoy, Ji and Lu assumed that the surface terms vanish for all the Mellin moments, and therefore implicitly assumed that the superpotential relating Belinfante and kinetic AM operators does not affect the $x$-dependence. In conclusion, the interpretation of the RHS of Eq.~\eqref{kindens} as the density of quark kinetic OAM should be considered as an \emph{ad hoc} definition rather than following from a natural operator definition.

The alternative derivation of the Ji relation for quarks in the rest frame proposed by Burkardt~\cite{Burkardt:2005hp} suggested that the integrand $\tfrac{x}{2}\left[H^q(x,0,0)+E^q(x,0,0)\right]$ could naturally be identified with an angular momentum decomposition of the transverse component of the quark angular momentum in a transversely polarized target. Ji, Xiong and Yuan tried to justify further this partonic interpretation by considering a slight variation of Burkardt's derivation~\cite{Ji:2012sj} and by extending it to gluons~\cite{Ji:2012ba}. However, a careful inspection reveals that this interpretation is not well founded. Indeed, the simple partonic interpretation proposed by the authors is based on the ``leading-twist'' part $\int\ud^2r_\perp\, \uvec r_\perp T^{++}$ of the Lorentz transformation generator $\int\ud^2r_\perp\, M^{++\perp}$. Lorentz invariance is then invoked to include the ``higher-twist'' part $\int\ud^2r_\perp r^+ T^{+\perp}$. While this argument is perfectly valid for the $x$-integrated quantities, it manifestly cannot be applied at the density level. Moreover, as pointed out in Refs.~\cite{Leader:2012md,Harindranath:2012wn}, Ji, Xiong and Yuan incorrectly interpreted $M^{++\perp}$ as the transverse angular momentum operator. In light-front quantization, $M^{++\perp}$ corresponds actually to the transverse boost operator which is kinematical and therefore has a simple partonic interpretation, unlike the transverse rotation operator $M^{+-\perp}$ which is dynamical and hence complicated due to the presence of interaction terms~\cite{Brodsky:1997de}. In order to obtain a boost-invariant sum rule, Ji, Xiong and Yuan~\cite{Ji:2012ba} focused on the transverse components of the Pauli-Lubanski vector and therefore included the higher-twist contribution $M^{+-\perp}$, but the strategy remained the same: do away with the higher-twist contribution using Lorentz symmetry~\cite{Ji:2013tva}. Moreover, the contribution associated with the $\bar C^{q,G}(0)$ energy-momentum form factor has been arbitrarily discarded by Ji, Xiong and Yuan~\cite{Ji:2012ba,Ji:2013tva}, based on the argument that it is higher-twist and does not contribute to the total Pauli-Lubanski vector. This contribution is however important once one considers quarks and gluons separately~\cite{Leader:2012ar,Hatta:2012jm,Harindranath:2013goa,Leader:2013jra}. Moreover, energy-momentum form factors being Lorentz scalars, naive twist counting of amplitudes does not provide any information about their {magnitudes and therefore does not justify neglecting $\bar C^{q,G}(0)$ in front of the other energy-momentum form factors. 

Clearly, the interpretation of the integrand of the Ji relation~\eqref{Jirel} as the density of parton AM is not justified. This is actually expected since the presence of the covariant derivative in the kinetic definition of OAM necessarily implies the contribution of higher-twist parton distributions. H\"agler, Mukherjee and Sch\"afer~\cite{Hagler:2003jw} proposed the following density of OAM involving both twist-2 and twist-3 GPDs
\begin{equation}\label{HMS}
\begin{split}
\langle L^q_\text{kin}\rangle(x)&=x\left[H^q(x,0,0)+E^q(x,0,0)\right.\\
&\qquad\left.+G^q_2(x,0,0)-2G^q_4(x,0,0)\right]-\Delta q(x)
\end{split}
\end{equation}
which when combined with Eq.~\eqref{kindens} led them to another form
\begin{equation}
\langle L^q_\text{kin}\rangle(x)=-x\left[G^q_2(x,0,0)-2G^q_4(x,0,0)\right].
\end{equation}
The interpretation of Eq.~\eqref{HMS} as the distribution of OAM has however been obtained within the Wandzura-Wilczek approximation, where the distinction between canonical and kinetic OAM disappears. Ji, Xiong and Yuan~\cite{Ji:2012ba} defined the densities of quark/gluon kinetic and potential OAM \emph{via} moments of the form~\eqref{qkinOAMmom} and showed how the latter are related to $D$ and $F$-type twist-3 GPDs. One has however to keep in mind that the gluon contributions defined in this way are those entering the JXY decomposition~\eqref{JXYdecomp} and not the standard kinetic decomposition. A more detailed discussion of the connection between twist-3 GPDs and densities of AM has been given by Hatta and Yoshida~\cite{Hatta:2012cs}. They stressed in particular that the density of kinetic OAM is ambiguous because $D$-type twist-3 GPDs involve two fractions of longitudinal momentum $x_1$ and $x_2$. This ambiguity does not show up for the density of canonical OAM since the latter has support only at $x_1=x_2$. This boils down to the fact that, contrary to ordinary derivatives,  covariant derivatives do not commute and therefore do not admit a unique non-local generalization~\cite{Lorce:2012ce}.

Motivated by the simple partonic picture behind the canonical AM decomposition of Jaffe and Manohar~\cite{Jaffe:1989jz}, Harindranath and Kundu~\cite{Harindranath:1998ve} defined a natural density of quark and gluon canonical OAM in the light-front gauge. Bashinsky and Jaffe~\cite{Bashinsky:1998if} then improved this definition by providing an expression invariant under residual gauge transformations. The density of parton OAM can in fact be defined in phase space and is easily expressed in terms of gauge-invariant phase-space distributions~\cite{Lorce:2011kd,Lorce:2011ni,Lorce:2012ce}
\begin{equation}
\langle L^{q,G}_{\mathcal W}\rangle(x,\uvec k_\perp,\uvec b_\perp)=(\uvec b_\perp\times\uvec k_\perp)_z\,\rho^{q,G}_{++}(x,\uvec k_\perp,\uvec b_\perp;\mathcal W).
\end{equation}
Like in the integrated version~\eqref{OAMWigner}, a staple-like Wilson line leads to the canonical version of the quark/gluon OAM, but this time it depends on whether the staple points toward the future $\mathcal W_{\sqsupset}$ or the past $\mathcal W_{\sqsubset}$. Indeed, based on the constraints imposed by parity and time-reversal symmetries, the relativistic phase-space distribution of unpolarized partons inside a longitudinally polarized nucleon can be decomposed into four modes
\begin{equation}
\begin{aligned}
\rho^{q,G}_{++}&=\rho^{q,G}_1+(\uvec b_\perp\cdot\uvec k_\perp)\,\rho^{q,G}_2+(\uvec b_\perp\times\uvec k_\perp)_z\,\rho^{q,G}_3\\
&\quad+(\uvec b_\perp\times\uvec k_\perp)_z\,(\uvec b_\perp\cdot\uvec k_\perp)\,\rho^{q,G}_4,
\end{aligned}
\end{equation}
where $\rho^{q,G}_i\equiv\rho^{q,G}_i(x,\uvec k^2_\perp,(\uvec b_\perp\cdot\uvec k_\perp)^2,\uvec b^2_\perp;\mathcal W)$. The functions $\rho_1$ and $\rho_2$ are associated with the real and imaginary parts of the GTMD $F_{11}$, whereas $\rho_3$ and $\rho_4$ are associated with the real and imaginary parts of the GTMD $F_{14}$. The factor $(\uvec b_\perp\cdot\uvec k_\perp)$ indicates that $\rho_2$ and $\rho_4$ are naive $\mathsf T$-odd, \emph{i.e.} change sign under $\mathcal W_{\sqsupset}\leftrightarrow\mathcal W_{\sqsubset}$. Note also that the only contribution to the canonical OAM surviving integration over $\uvec b_\perp$ or $\uvec k_\perp$ is $\rho_3$. For a straight Wilson line $\mathcal W_\mid$, one obtains the density of JXY quark and gluon OAM~\cite{Ji:2012sj,Ji:2012ba} provided that one integrates at least over $\uvec k_\perp$~\cite{Lorce:2012ce}.
\newline

We close this section with a discussion of the Ji relation~\eqref{Jirel} in position space. Since the $t$ dependence of twist-2 GPDs contains the information about the spatial distribution of partons~\cite{Burkardt:2000za,Burkardt:2002hr}, it has been suggested by Polyakov~\cite{Polyakov:2002yz,Goeke:2007fp} that the Ji relation generalized to $t\neq 0$ contains the information about the spatial distribution of Belinfante AM
\begin{equation}
\langle J^{q,G}_\text{kin}\rangle(t)=\tfrac{1}{2}\int\ud x\,x[H^{q,G}(x,0,t)+E^{q,G}(x,0,t)].
\end{equation}
Within the scalar diquark model with Pauli-Villars regularization where $\langle L^q_\text{kin}\rangle=\langle L^q_\text{can}\rangle$, Adhikari and Burkardt~\cite{Adhikari:2013ima} observed that the quark kinetic and canonical spatial densities do not coincide $\langle L^q_\text{kin}\rangle(\uvec b_\perp)\neq\langle L^q_\text{can}\rangle(\uvec b_\perp)$. Note however that these spatial densities have been defined in the Drell-Yan frame as simple two-dimensional Fourier transform $\langle L^q\rangle(\uvec b_\perp)=\int\frac{\ud^2\Delta_\perp}{(2\pi)^2}\,e^{-i\uvec \Delta_\perp\cdot\uvec b_\perp}\,\langle L^q\rangle(t=-\uvec\Delta^2_\perp)$. While this is justified for the canonical OAM and helicity distributions, the actual spatial density of total AM is given by the Fourier transform\footnote{In the non-relativistic interpretation based on the large-$N_c$ picture used by Polyakov, the spatial distribution is given by the three-dimensional Fourier transform of $\langle J^q_\text{kin}\rangle(t)+\tfrac{2t}{3}\,\tfrac{\partial}{\partial t}\langle J^q_\text{kin}\rangle(t)$~\cite{Polyakov:2002yz}. Note that a quadrupole contribution has implicitly been discarded in Ref.~\cite{Goeke:2007fp} by doing the substitution $-\uvec\Delta^2_\perp\mapsto\tfrac{2t}{3}$.} of $\langle J^q_\text{kin}\rangle(t)+t\,\tfrac{\partial}{\partial t}\langle J^q_\text{kin}\rangle(t)$~\cite{Leader:2013jra}. The second term does not contribute to the total AM since the integral over all space amounts to set $t=0$. Note also that the spatial density of quark kinetic OAM has been defined as $\langle L^q_\text{kin}\rangle(\uvec b_\perp)=\langle J^q_\text{kin}\rangle (\uvec b_\perp)-\langle S^q\rangle (\uvec b_\perp)$, but this relation is actually spoiled by the superpotential term relating Belinfante and kinetic forms of the energy-momentum tensor~\cite{Leader:2013jra}. Further investigations are once more needed.

\section{The orbital angular momentum from the lattice}\label{sec5}

   As mentioned in the Introduction, the quark spin as determined from DIS experiments is about
25\% of the proton spin and recent global analyses~\cite{deFlorian:2014yva,Nocera:2014gqa} including high-statistics 2009 STAR~\cite{Adamczyk:2014ozi} and PHENIX~\cite{Adare:2014hsq} data show an evidence of non-zero gluon helicity in the proton. For $Q^2=10$ ${\rm GeV}^2$, it is found that the gluon helicity distribution $\Delta G(x,Q^2)$ is positive and different from zero in the momentum fraction range $0.05\leq x \leq0.2$
and the integral $\int_{0.05}^1 \ud x \,\Delta G(x)$ is  $ 0.17 \pm 0.06$ -- about 35\% of the proton 
spin~\cite{Nocera:2014gqa}. However, the results have very large uncertainty in the small 
$x$-region~\cite{deFlorian:2014yva,Nocera:2014gqa}. Since the sum of the quark and gluon spin do not saturate
the total proton spin, it leaves quite a bit of room for the OAM of quarks and gluons.

Now that we have discussed the decomposition of the proton spin in terms of the quark and gluon contributions
in Eqs.~\eqref{chen_decomp} and \eqref{ji_decomp}, and how to access OAM using GPDs 
and  GTMDs, it would be useful and imperative to calculate them within lattice QCD which consists in {\it ab initio} calculations with controllable statistical and systematic errors. 

\subsection{Mellin moment approach}  \label{mma}

We shall first consider Ji's decomposition~\eqref{ji_decomp} which involves the quark spin, quark kinetic OAM, and gluon total AM.  Since the decomposition of longitudinal AM is boost-invariant, one can calculate the Mellin moments of the GPDs $H$ and $E$ through the matrix elements of the energy-momentun tensor (EMT) near the rest frame like in the case of the parton structure functions. The lattice calculation of physical quantities involving small lattice momenta near the lowest non-zero lattice momentum, \emph{i.e.} $pa \sim \frac{2\pi}{La}$, with $a$ the lattice spacing and $La$ the physical lattice size, are less noisy than those involving $pa  \gg \frac{2\pi}{La}$. 
    
The Lorentz group generator $J^{\mu\nu}$ is given by~\cite{Weingerg:1972}
\begin{equation}
 J^{\mu\nu} =\int \ud^3 r \,M^{0\mu\nu} (r) .
\label{ang_mom_generator}
\end{equation}
Here $M^{0\mu\nu}$ is the Lorentz group generator density which is defined as
\begin{equation}
  M^{\mu\nu\alpha} (r) 
  = {\mathcal T}^{\mu\alpha} r^\nu - {\mathcal T}^{\mu\nu} r^\alpha ,
\label{ang_mom_density}
\end{equation}
where $\mathcal{T}^{\mu\nu}$ is the Belinfante-improved EMT which is gauge invariant, symmetric $\mathcal{T}^{\nu\mu}=\mathcal{T}^{\mu\nu}$ and conserved 
$\partial_{\mu} \mathcal{T}^{\mu\nu}=0$~\cite{Treiman:1986ep}. This EMT can be decomposed into quark and gluon kinetic contributions
\begin{equation}
  \mathcal{T}^{\mu\nu} 
  = \mathcal{T}^{\mu\nu}_{\text{kin},q} + \mathcal{T}^{\mu\nu}_{\text{kin},G} ,
  \label{energy_mom_split}
\end{equation}
where $\mathcal{T}^{\mu\nu}_{\text{kin},q}$ and $\mathcal{T}^{\mu\nu}_{\text{kin},G}$ have the following 
form
\begin{align}
 {\mathcal T}^{\mu\nu}_{\text{kin},q} 
  &= \tfrac{i}{4}\sum_f \overline {\psi}_f 
    [\gamma^{\mu}\! \!\stackrel{\leftrightarrow}{D}\!\!\!\!\!\!\!\phantom{D}^{\nu}
     +\gamma^{\nu}\! \!\stackrel{\leftrightarrow}{D}\!\!\!\!\!\!\!\phantom{D}^{\mu}]  \psi_f,
  \label{q_contrib_def_1}\\
{\mathcal T}^{\mu\nu}_{\text{kin},G} 
  &=  - 2\,\uTr[F^{\mu\lambda} F^{\nu}_{\phantom{\nu}\lambda}]+ \tfrac{1}{2}\, g^{\mu\nu}\,\uTr[F^2]. 
  \label{g_contrib_def_1}
\end{align}
with $\stackrel{\leftrightarrow}{D}\!\!\!\!\!\!\!\phantom{D}^{\nu}=\stackrel{\rightarrow}{D}\!\!\!\!\!\!\!\phantom{D}^{\nu}  - \stackrel{\leftarrow} {D}\!\!\!\!\!\!\!\phantom{D}^{\nu}$.
The AM operator is defined from the Lorentz group generator $J^{\mu\nu}$ in Eq.~\eqref{ang_mom_generator} as
\begin{equation}
 J^i = \tfrac{1}{2}\,\epsilon^{ijk} \int \ud^3 r\, M^{0jk} (r) .
\label{ang_mom_operator}
\end{equation}   

Since the quark kinetic OAM $\uvec L^q_\text{kin}$ in Eq.~\eqref{dec5} and the gluon total AM operators $\uvec J^G_\text{kin}$ in Eq.~\eqref{JG} depend on
the spatial vector $\uvec{r}$, a straightforward 
%application of the 
lattice calculation of the operators with
forward matrix elements are complicated 
by the periodic condition of the lattice, and may lead to wrong results~\cite{Wilcox:2002zt}. Instead of calculating $\uvec L^q_\text{kin}$ and $\uvec J^G_\text{kin}$ directly, they can be obtained from the energy-momentum form factors in the nucleon which parametrize off-forward matrix elements of the EMT~\cite{Ji:1996ek,Bakker:2004ib,Lorce:2015lna}.
\newline

To carry out lattice calculations, one works with the Monte-Carlo ensembles of the 2-
  and 3-point path-integral correlation functions in the Euclidean space.      
   The Euclidean energy-momentum operators for quarks and gluons are~\cite{Deka:2008xr}
\begin{align}
\label{eq:q_contrib_def_2}
{\mathcal T}_{4i,q}^{E} 
  &=  -\tfrac{i}{4}\sum_f \overline {\psi}_f 
  [ 
    \gamma^E_4 \!\!\stackrel{\leftrightarrow} D\!\!\!\!\!\!\!\phantom{D}^E_i 
    + \gamma^E_i \!\!\stackrel{\leftrightarrow} D\!\!\!\!\!\!\!\phantom{D}^E_4
   ] \psi_f , \\
\label{eq:g_contrib_def_2}
 {\mathcal T}_{4i,G}^{E} 
  &= 2i\uTr[F^E_{4\alpha} F^E_{i\alpha} ],
\end{align}
where the Euclidean $\gamma^E_{\mu}$ matrices satisfy $\{\gamma^E_{\mu}, \gamma^E_{\nu}\} = 2 \delta_{\mu\nu}$
and the Euclidean components of a four-vector are defined as $p^E_{\mu} = (\vec{p}, i p^0)$.
% The gauge field tensor $F_{\mu\nu}$ are those in the Euclidean space. 
In the following,  the Pauli-Sakurai representation for the gamma matrices~\cite{Montvay:1994cy,Best:1997qp} is used and the covariant
derivative is the point-split lattice operator involving the gauge link $U_{\mu}$~\cite{Kronfeld:1984zv}. 
%\green{From now on, we will drop for convenience the label $E$ on Euclidean quantities.}

The form factors of the quark and gluon EMTs are defined as 
\begin{equation}
\langle p', S'  | {\mathcal T}_{4i,a}^{E} | p, S\rangle=\tfrac{1}{2}\, \overline{u}^{E}(p',S')\Gamma^E_{4i,a} u^{E}(p,S) \label{mat_element_2}
\end{equation}
with $a=q,G$ and\footnote{Note that the energy-momentum form factor $T^a_4$ does not contribute to \eqref{mat_element_3} as it is associated with the structure $\delta_{4i}=0$.} 
\begin{equation}
\begin{aligned}
\Gamma^E_{4i,a}&= (\gamma_4 P_i    +  \gamma_i P_4)\, T^a_1(-\Delta^2) \\
&\quad- \frac{(P_4 \sigma_{i\alpha} +  P_i \sigma_{4\alpha}) \Delta_{\alpha}}{2m}\,T^a_2(-\Delta^2)\\
&\quad- \frac{i\Delta_4 \Delta_i}{m}\,T^a_3(-\Delta^2),\label{mat_element_3}
\end{aligned}
\end{equation}
where $P=\tfrac{p'+p}{2}$, $\Delta=p'-p$, 
% Comment: we don't need this definition -- $S^\mu$ is the covariant spin vector 
and the normalization conditions for the nucleon spinors are
\begin{align}
\bar{u}^{E}(p,S)\, u^{E}(p,S) &= 1\ \nonumber \\
\sum_S  u^{E}(p,S)\, \bar{u}^{E}(p,S)&=\frac{\slashed{p} + M}{2M}.
\label{norm_cond_euc}
\end{align}
The EMT form factors 
$T_1$ and $T_2$ resemble the Dirac and Pauli form factors $F_1$ and $F_2$ associated with the
electromagnetic current. While the nucleon electric charge and magnetic dipole moment respectively equal $F_1$ 
and $F_1 + F_2$ at $\Delta^2=0$, the quark and gluon kinetic momenta and total kinetic AM fraction are shown to
be respectively equal to $T^a_1$ and $T^a_1+T^a_2$ at $\Delta^2=0$~\cite{Ji:1996ek}.
\newline

Lattice theory has been able to  accommodate chiral symmetry of the vector gauge theories finally, the lack 
of which has hampered the development of chiral fermions on the lattice for many years. It is shown that when 
the lattice massless Dirac operator satisfies the Gingparg-Wilson relation 
$\gamma^E_5 D^E + D^E \gamma^E_5 = a D^E\gamma^E_5D^E$
with the overlap fermion being an explicit example~\cite{Neuberger:1997fp}, the
modified chiral transformation leaves the action invariant and gives rise to a chiral Jacobian factor
$J = e^{-2i\alpha \uTr[\gamma^E_5 (1 - \tfrac{1}{2}\,a D^E)]}$ from the fermion determinant~\cite{Luscher:1998pqa}.
The index theorem~\cite{Hasenfratz:1998ri} shows that this Jacobian factor carries the correct chiral anomaly.
Thus, the Atiya-Singer theorem is satisfied on the lattice with a smooth gauge configuration, \emph{i.e.} the total topological 
charge equals the index (number of left-handed zero modes minus right-handed ones) of the configuration. Further,
it is shown that the local version of the overlap Dirac operator gives the topological
charge density operator in the continuum~\cite{Kikukawa:1998pd,Adams:1998eg,Fujikawa:1998if,Suzuki:1998yz}, namely
\begin{equation}  \label{top_q}
\uTr[\gamma^E_5 (1 - \tfrac{1}{2}\,a D^E_{ov}(r,r) )]= \tfrac{1}{8 \pi^2} \uTr[F^E_{\mu\nu} 
\tilde{F}^{E\mu\nu}]+ \mathcal{O}(a).
\end{equation} 
This definition is superior to previous lattice constructs with gauge links. Since the overlap operator
is exponentially local, the topological charge operator so defined is chirally smoothed with 
neighboring links. This operator has been used to study the topological properties of the vacuum, and
charged coherent membranes with lower-dimensional long-range order~\cite{Horvath:2003yj,Horvath:2005cv} were discovered which have implications on chiral symmetry breaking. 
 
In view of the fact that the overlap operator definition of the topological charge has good properties, 
 a similar connection between  $F^E_{\mu\nu}$ and the overlap Dirac operator has been 
derived~\cite{Liu:2007hq,Alexandru:2008fu} 
\begin{equation}
  \uTr_s [\sigma^E_{\mu\nu} D^E_{\rmsmall{ov}}(r,r)] 
  =  c_T\, a^2 F^E_{\mu\nu}(r) + {\mathcal O}(a^3),
\label{Glue_Tensor}
\end{equation}
where $\uTr_s$ is the trace over spin and $c_T = 0.11157$ is the proportionality constant 
in the continuum limit for the parameter $\kappa = 0.19$ in the Wilson kernel of the overlap 
operator which is used in the lattice calculation~\cite{Liu:2007hq}. The overlap Dirac operator $D^E_{ov}(r,r')$ is exponentially
local and the gauge field $F^E_{\mu\nu}$ as defined in Eq.~\eqref{Glue_Tensor} is chirally smoothed
so that it admits good signals for the gluon momentum and total AM in the lattice 
calculation~\cite{Deka:2013zha}. This method proved to be much less noisy than the standrad approach based on the clover-leaf definition from the gauge links~\cite{Mandula:1982us}.

\subsubsection{Sum rules and renormalization}   \label{sumrules}

\hspace{0.5cm}
The momentum and total AM contributions from quarks and gluons depend 
on the renormalization scale and scheme individually, but their sums do not
because the total momentum and AM of the nucleon are conserved. One can use these
sum rules as the renormalization conditions on the lattice.

Substituting the EMT matrix elements in Eq.~\eqref{mat_element_2}
to the matrix elements which define the AM in Eq.~\eqref{ang_mom_operator} 
and similarly for the momentum, it is shown~\cite{Ji:1996ek} that
\begin{align}
  \label{ang_op_def_split_3}
  \langle J^{q,G}_\text{kin}\rangle &= Z_{q,G}^L \,\tfrac{1}{2}\, [T^{q,G}_1(0) + T^{q,G}_2(0)],\\
  \langle x\rangle_{q,G} &= Z_{q,G}^L\, T^{q,G}_1(0),
\label{momentum_fraction}
\end{align}
where $Z_{q,G}^L$ is the renormalization constant for the lattice quark/gluon operators, and
$\langle x\rangle_{q,G}$ is the second Mellin moment of the unpolarized
parton distribution function (PDF) $\int\ud x\,x\,f^{q,G}_1(x)$, which represents the (average) momentum 
fraction carried by quarks or gluons inside a nucleon. The other form factor, 
$T^{q,G}_2(0)$, can be interpreted as the anomalous gravitomagnetic moment 
in analogy to the anomalous magnetic moment $F_2(0)$~\cite{Teryaev:1999su,Brodsky:2000ii}.

Since momentum is always conserved and the nucleon has a total spin of 
$1/2$, we write the momentum and AM sum rules 
using 
Eqs.~\eqref{ang_op_def_split_3}~and~\eqref{momentum_fraction}, 
as
\begin{align}   \label{eq:mom_sum_rule}
Z_q^L \,T^q_1 (0) + Z_G^L \,T^G_1 (0)  &= 1,\\ 
 \tfrac{1}{2}\,\{   Z_q^L \,[ T^q_1 (0) + T^q_2 (0)]+ Z_G^L \,[ T^G_1 (0)  + T^G_2 (0)] \} & =\tfrac{1}{2}.
\label{eq:ang_mom_sum_rule}
 \end{align}
It is interesting to note that from Eqs.~\eqref{eq:mom_sum_rule} and 
\eqref{eq:ang_mom_sum_rule}, one obtains that the sum of the $T_2(0)$'s for the 
quarks and gluons is zero, \emph{i.e.}
\begin{equation}
Z_q^L \,T^q_2 (0) + Z_G ^L\, T^G_2 (0) = 0.
\label{eq:T_2_sum}
\end{equation}
The vanishing of the total anomalous gravitomagnetic moment ($T_2 (0)=0)$ in the context of a spin-$1/2$ particle 
was first derived classically from 
the post-Newtonian manifestation of equivalence principle~\cite{Kobzarev:1962wt}. More 
recently, this has been proven by Brodsky {\it et al.}~\cite{Brodsky:2000ii} for composite 
systems from the light-front Fock space representation.

One can use these sum rules and the raw lattice results to obtain the lattice renormalization
constants $Z_q^L$ and $Z_G^L$, and then use perturbation theory~\cite{Glatzmaier:2014sya} to calculate the quark-gluon
mixing and renormalization in order to match to the $\overline{MS}$ scheme at 2 GeV which preserves the sum rules. 

\subsubsection{Results of lattice calculations}   \label{lattice_results}
\begin{figure}[t]
\centering
\subfigure[]
{\rotatebox{0}{\includegraphics[width=0.4\hsize]{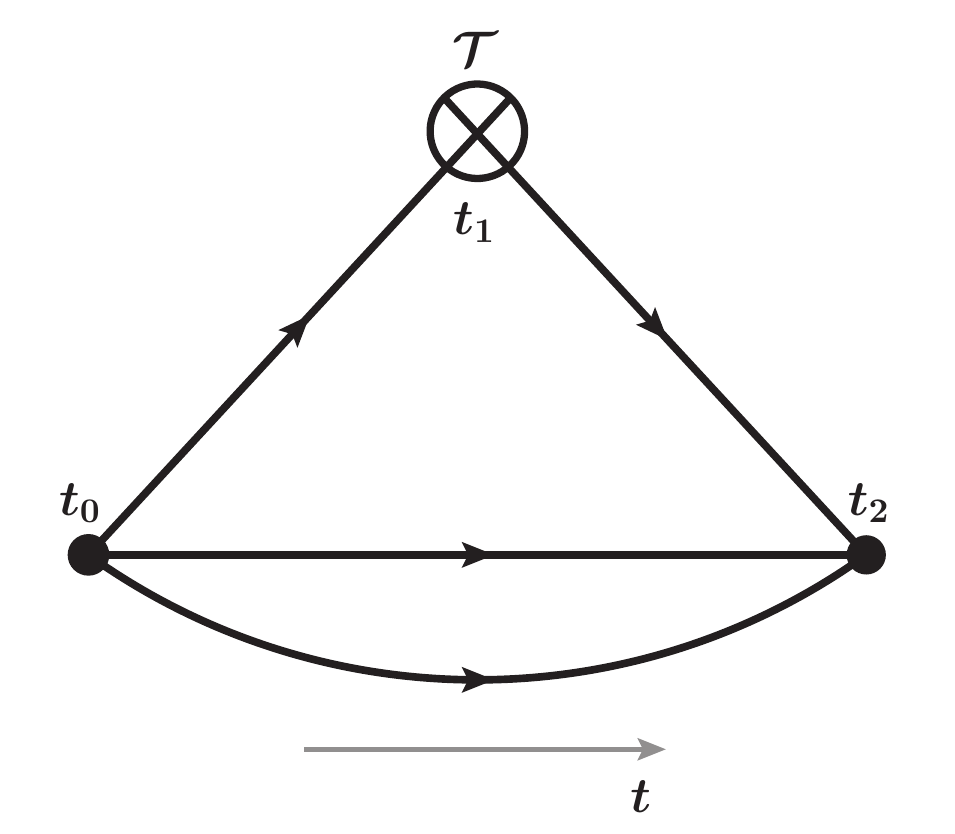}}
\label{fig:connected_insertion}}
\hspace{2mm}
\subfigure[]
{\rotatebox{0}{\includegraphics[width=0.4\hsize]{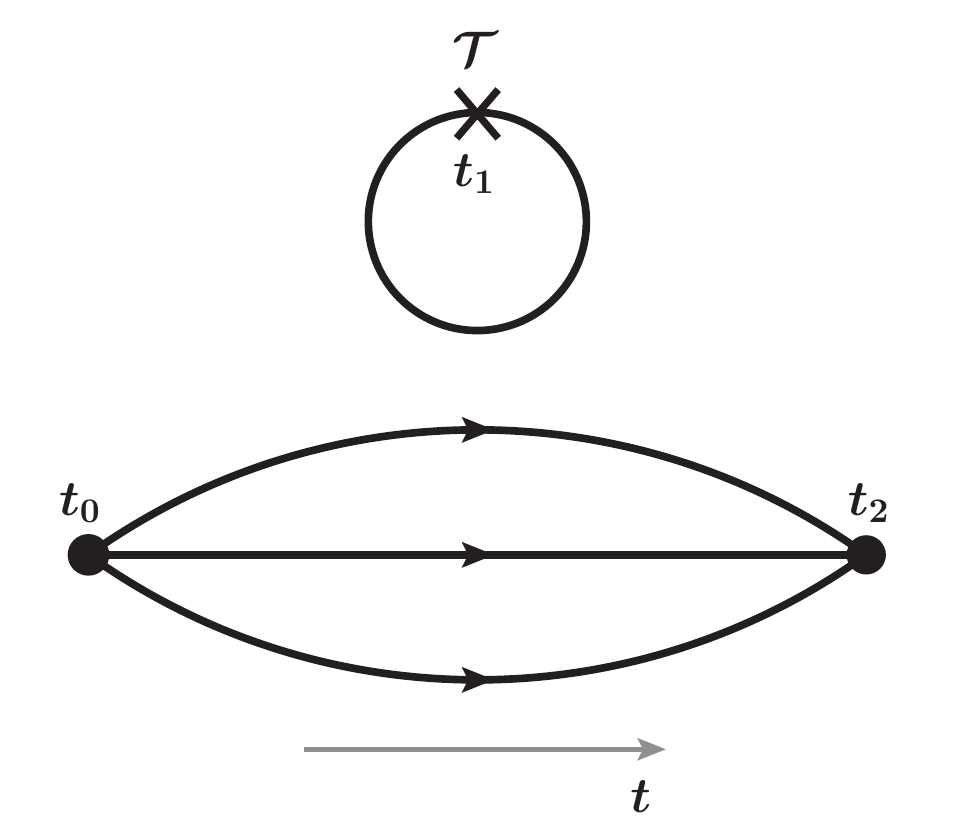}}
 \label{fig:disconnected_insertion}}
\caption{Quark line diagrams of the 3-point function with current insertion in the Euclidean
         path integral formalism. 
         (a) Connected insertions (CI), and
         (b) disconnected insertions (DI).}
\label{fig:ci_and_di}
\end{figure}

\hspace{0.5cm}
Before presenting the lattice results, we should point out that
the 3-point functions for quarks which are needed to extract the form factors
in Eq.~\eqref{mat_element_3} have two topologically distinct contributions in the
path-integral diagrams: one from connected insertions (CI) and the other from 
disconnected insertions (DI)~\cite{Liu:1993cv,Liu:1998um,Liu:1999ak,Liu:2012ch}, see 
Fig.~\ref{fig:ci_and_di}. They arise in different Wick contractions, and 
it needs to be stressed that they are not Feynman diagrams in perturbation theory. In 
the case of CI, quark/antiquark fields from the operator are contracted with the 
quark/antiquark fields of the proton interpolating fields. It represents the valence and
the higher Fock space contributions from the $Z$-graphs. In the case of DI, the 
quark/antiquark fields from the operator contract themselves to form a current loop, which
represents the vacuum polarization of the disconnected sea quarks. Also, although the quarks lines in the loop and the nucleon propagator appear
to be disconnected in Fig~\ref{fig:ci_and_di}(b), they are in fact correlated
through the gauge background fluctuations. In practice, the uncorrelated part of
the loop and the proton propagator is subtracted. For the nucleon, the up and down quarks contribute to both CI and DI, while the 
strange and charm quarks contribute to the DI only.

As for the quark OAM, lattice calculations have been carried out for the CI~\cite{Mathur:1999uf,Hagler:2003jd,Gockeler:2003jfa,Brommel:2007sb,Bratt:2010jn,Alexandrou:2011nr,Syritsyn:2011vk,Alexandrou:2013joa}. They are obtained by subtracting 
the quark spin contributions from those of the quark kinetic total AM~\eqref{JiOAMrel}. It has been shown 
that the contributions from $u$ and $d$ quarks almost cancel each other. Thus for CI, quark OAM turns out to be small in the quenched 
calculations~\cite{Mathur:1999uf,Gockeler:2003jfa} and nearly zero in dynamical fermion 
calculations~\cite{Brommel:2007sb,Bratt:2010jn,Alexandrou:2011nr,Syritsyn:2011vk,Alexandrou:2013joa}.

The first complete decomposition of proton spin into quark spin, quark OAM and
 gluon total AM together with quark and gluon momenta has been carried out by Deka {\it et al.} 
($\chi QCD$ Collaboration)~\cite{Deka:2013zha}. The calculation is done on a quenched lattice
with 3 valence quark masses and extrapolated to the physical pion mass. The results are presented in Table~\ref{tab:chiral} and
 Fig.~\ref{fig:pie_diag}. 

\begin{table*}[t]
  \centering
  \caption{Renormalized results for connected (CI) and disconnected (DI) insertions in $\overline{MS}$ scheme at $\mu = 2$~GeV.}
 \begin{tabular}{c|cc|cccc}
    \hline\noalign{\smallskip}
    &  CI($u$) &  CI($d$)  &  CI($u+d$) &   DI($u$/$d$) &  DI($s$) &  Gluons \\
    \noalign{\smallskip}\hline\noalign{\smallskip}
    $\langle x \rangle$
    & 0.413(38)  &  0.150(19) & 0.565(43) & 0.038(7) & 0.024(6) & 0.334(55) \\
    $T_2(0)$ 
    & 0.286(108)  & -0.220(77) & 0.062(21) & -0.002(2) & -0.001(3) & -0.056(51) \\
    $2\langle J_\text{kin}\rangle$ 
    &  0.700(123)  & -0.069(79) & 0.628(49) & 0.036(7) & 0.023(7) & 0.278(75)\\
    $g_A$
    &  0.91(11)  & -0.30(12)   & 0.62(9)  &  -0.12(1)  &  -0.12(1) & \--- \\
    $2\langle L_\text{kin}\rangle$
    &  -0.21(16)    &  0.23(15)   &  0.01(10)  &  0.16(1)  &  0.14(1) & \--- \\
   \noalign{\smallskip}\hline
  \end{tabular}
\label{tab:chiral}
 \end{table*}
%\Blindtext\Blindtext  
%
%
\begin{figure}[t]
  \centering
  \subfigure[]
  {\rotatebox{0}%
    {\includegraphics[width=0.485\textwidth]{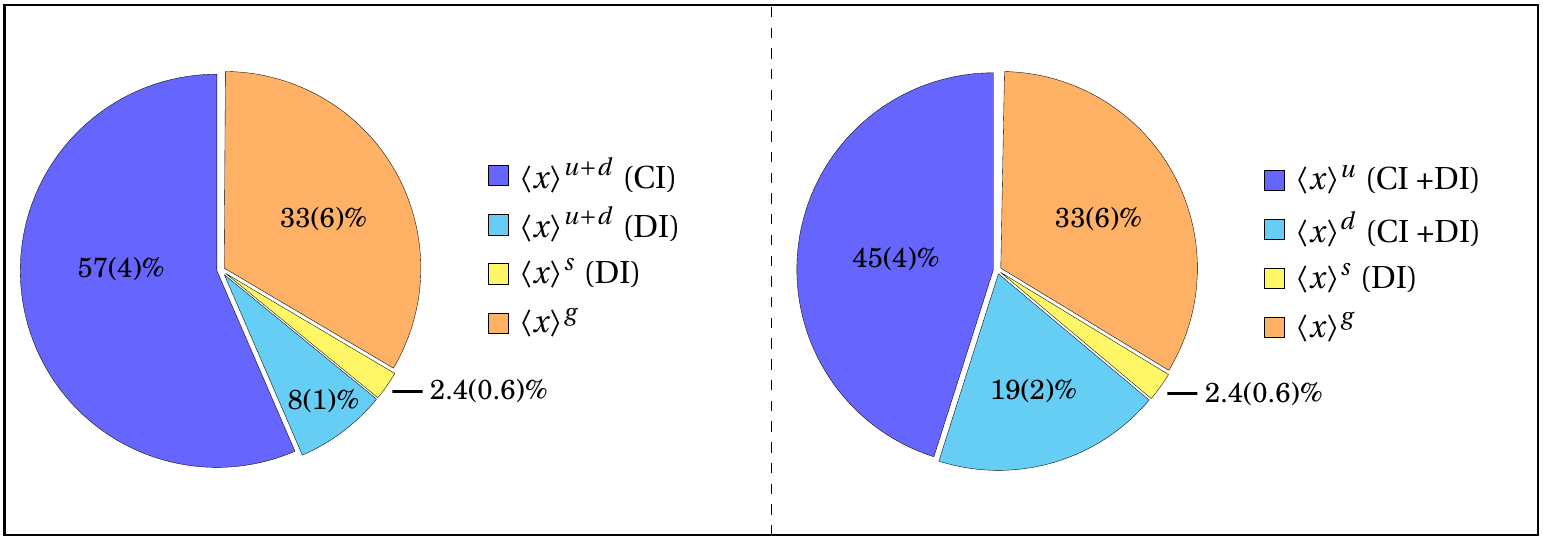}}
    \label{fig:pie_diag_first_mom}
  }
  \subfigure[]
  {\rotatebox{0}%
   {\includegraphics[width=0.485\textwidth]{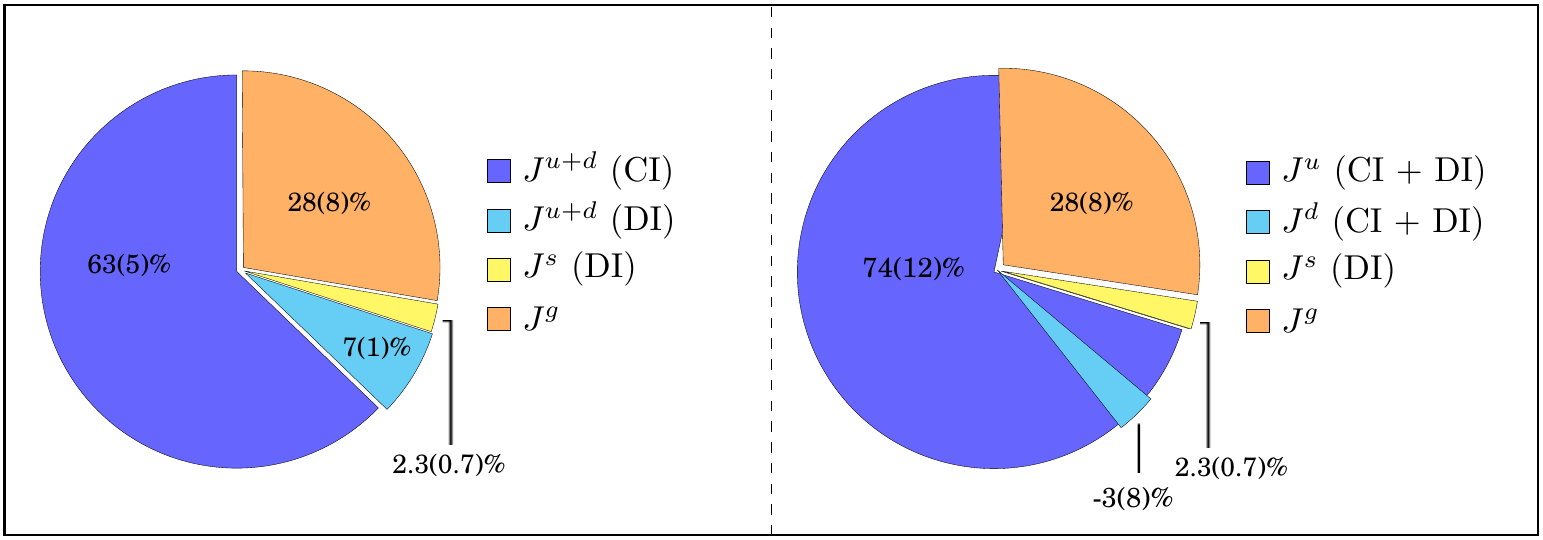}}
   \label{fig:pie_diag_am}
  }
  \subfigure[] 
  {\rotatebox{0}%
    {\includegraphics[width=0.485\textwidth]{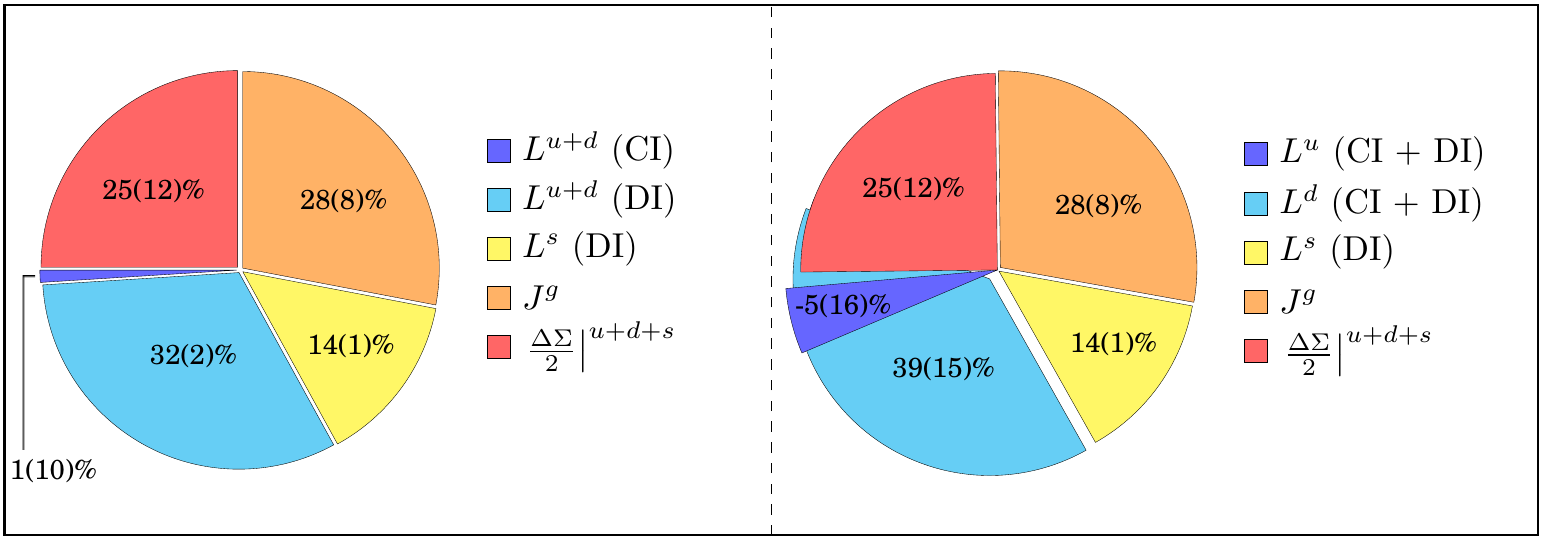}}
    \label{fig:pie_diag_orb_am}
  }
  \caption{Pie charts for the quark and gluon\
    (a)\ momentum fractions,\ 
    (b)\ AM fractions,\ 
    and\
    (c)\ spin and OAM fractions.\ 
    The left panels show the quark contributions separately for CI and DI, and the right panels 
    show the quark contributions for each flavor with CI and DI summed together for $u$ and 
    $d$ quarks.%
  }
  \label{fig:pie_diag}
\end{figure}
We see from Table~\ref{tab:chiral} that the strange momentum fraction 
$\langle x\rangle_s = 0.024(6)$ is in the range of uncertainty of $\langle x\rangle_s$ from 
the CTEQ fitting of the PDF from experiments
$0.018 < \langle x\rangle_s < 0.040$~\cite{Lai:2007np}. The gluon momentum fraction of 
$0.334 (55)$ is smaller than, say, the CTEQ4M fit of $0.42$ at 
$\mu = 1.6$~GeV~\cite{Lai:1996mg}, but only by $1.5 \sigma$. The smallness of this
value of $\langle x\rangle_G$ in comparison to the experiment could be in part due to the 
fact that this is a quenched calculation. It is expected that the gluon momentum fraction will be larger 
than the present result when dynamical configurations with light fermions are used in the 
calculation. 

For the unrenormalized lattice results, $[T_2^u(0) + T_2^d(0)]_\text{CI}$ is found positive and $T_{2}^G (0)$ negative, 
so that including the small $[T_2^u (0) + T_2^d (0) + T_2^s (0)]_\text{DI}$, the total sum
can be naturally constrained to be zero, see Eq.~\eqref{eq:T_2_sum} with the lattice normalization 
constants $Z_q^L = 1.05$ and $Z_G^L = 1.05$ close to unity. As discussed earlier, the vanishing of  
the total $T_2(0)$ is a consequence of momentum and AM conservation. 

The flavor-singlet $g_A^0=g^u_A+g^d_A+g^s_A=\Delta\Sigma=2\langle S^q\rangle$, which represents the quark spin fraction to the nucleon spin, has been 
calculated before on the same lattice~\cite{Dong:1995rx}. The quark spin contribution are subtracted in Table~\ref{tab:chiral} 
and Fig.~\ref{fig:pie_diag} from the quark total AM contributions $\langle J^q_\text{kin}\rangle$ to obtain the quark OAM contributions $\langle L^q_\text{kin}\rangle$. As we see in Table~\ref{tab:chiral}, the OAM fractions $2\langle L^q_\text{kin}\rangle$ for the $u$ and $d$ quarks in the CI have different signs and add up to $0.01(10)$, \emph{i.e.}
essentially zero. This is the same pattern which has been seen with dynamical fermion configurations and light quarks, as pointed out earlier. The large OAM fractions $2\langle L^q_\text{kin}\rangle$ for the $u$/$d$ and $s$ quarks in the DI is due to the 
fact that $g_A^0$ in the DI is large and negative, about $-0.12(1)$ for each of the three 
flavors. All together, the quark OAM constitutes a fraction of 
$0.47(13)$ of the nucleon spin. The majority of it comes from the DI. 

As far as the spin decomposition is concerned, it is found that
the quark spin constitutes 25(12)\% of the proton spin, the gluon total AM takes 
28(8)\% and the rest is due to the quark kinetic OAM which is 47(13)\%. 
%The lattice
%quark and gluon EMT operators are renomalized with the momentum and
%AM sum rules. The matching from lattice to the $\overline{MS}$ scheme at 
%2 GeV is carried out perturbatively~\cite{Glatzmaier:2014sya}. 

Since this calculation is based on a quenched approximation which is known to contain
large uncontrolled systematic errors, it is essential to repeat this calculation with dynamical
fermions of light quarks and large physical volume. However, we expect that the quark OAM fraction may still be large in the dynamical calculation. First of all, the quark spin
fraction is about 20\% - 30\% from DIS and the gluon momentum fraction from CTEQ4M fit gives
$\langle x\rangle_G = 0.42$ at $\mu = 1.6$~GeV~\cite{Lai:1996mg} which is not far from our 
calculation at $\mu = 2$ GeV.  Since the gluon total AM is $\langle x\rangle_G + T^G_2(0)$, 
if we assume that $T^G_2(0)$ in the dynamical calculation is not too far from the small $T^G_2(0) = -0.056(51)$ obtained in the quenched approximation, the gluon total AM will be $\sim 36\%$. In this case, the quark 
OAM fraction will still be at the 40--50\% level. This will be checked by the upcoming lattice calculations with dynamical fermion configurations at physical pion mass. 

In the naive constituent quark model, the proton spin comes entirely from the quark spin. On the 
other hand, in the Skyrme model~\cite{Adkins:1983ya} the proton spin originates solely from the OAM of the collective rotational motion of the pion field~\cite{Li:1994zp}. What is found in the present lattice calculation seems to suggest that the QCD picture, aside from the gluon contribution, is somewhere in between 
these two models, indicating a large contribution of the quark OAM due to the meson cloud ($q\overline{q}$ pairs in the higher Fock space)  in the nucleon.

\subsubsection{Asymmetric energy-momentum tensor}

Let us stress that the standard Mellin moment approach is based on the Belinfante-improved EMT which is symmetric and therefore leads only to the total AM of quarks and gluons, see Eqs.~\eqref{ang_mom_generator} and  \eqref{ang_mom_density}. 
In the quark sector, one can alternatively use the asymmetric EMT
\begin{equation}
\widetilde{T}^{\mu\nu}_{\text{kin},q}= \tfrac{i}{2}\sum_f \overline {\psi}_f \gamma^{\mu}\! \!\stackrel{\leftrightarrow}{D}\!\!\!\!\!\!\!\phantom{D}^{\nu}
      \psi_f
\end{equation}
which differs from the Belinfante-improved EMT~\eqref{q_contrib_def_1} by a gauge-invariant antisymmetric superpotential term~\cite{Leader:2013jra}
\begin{equation}
\widetilde{T}^{\mu\nu}_{\text{kin},q}=\mathcal T^{\mu\nu}_{\text{kin},q}- \tfrac{1}{4}\,\partial_\alpha(\epsilon^{\mu\nu\alpha\beta}\sum_f\overline\psi_f\gamma_\beta\gamma_5\psi_f)
\end{equation}
provided that the QCD equations of motion $(i\,\uslash D+m_f)\psi_f=0$ hold. From this asymmetric EMT, one can directly compute the quark kinetic OAM
\begin{equation}\label{kinOAMrel}
L^i_{\text{kin},q}=\tfrac{1}{2}\,\epsilon^{ijk} \int \ud^3 r\, (\widetilde{T}^{0k}_{\text{kin},q}r^j
-\widetilde{T}^{0j}_{\text{kin},q}r^k).
\end{equation}
The parametrization of the off-forward matrix elements of the asymmetric EMT, beside the structures in Eq.~\eqref{mat_element_3}, involves a new structure $ (\gamma_4 P_i   -  \gamma_i P_4)\, T^q_5(-\Delta^2)$~\cite{Bakker:2004ib,Lorce:2015lna}. The quark spin and kinetic OAM fractions are then respectively given by $-T^q_5$ and $T^q_1+T^q_2+T^q_5$ at $\Delta^2=0$~\cite{Bakker:2004ib,Leader:2013jra,Lorce:2015lna}. To the best of our knowledge, this approach has not been investigated so far on the lattice. In particular, it would be interesting to compare the lattice calculation of $-T^q_5$ and $g^q_A$ since both represent the quark spin fraction at $\Delta^2=0$ obtained from two different operators.

\subsection{Large momentum approach}\label{LaMETsec}

To carry out lattice calculations of all the terms in the generic sum rule of Eq.~\eqref{decomposition},
one needs to define the fields $A^\mu_\pure$ and $A^\mu_\phys$ on the lattice.
Moreover, the partonic picture
of OAM is naturally depicted in the light-front formalism and related to GPDs and GTMDs as discussed in
Sec.~\ref{sec3}.  Unfortunately, the light-front coordinates are not accessible to lattice QCD calculation since the latter is based 
on Euclidean path-integral formulation. 

    To bridge the gap between the light-front formulation and the lattice calculation, it is 
suggested by Ji~\cite{Ji:2013dva} that light-front parton distributions can be accessed as the infinite momentum limit of 
equal-time correlators.  To do so, one calculates the equal-time correlators on the lattice at large but finite hadron momentum, 
called quasi parton distributions, and match them to the $\overline{MS}$ scheme at a certain scale 
$\mu$ in the continuum. After the usual continuum extrapolation of the lattice results, a large momentum effective field 
theory (LaMET)~\cite{Ji:2014gla}, which takes care of the non-commuting UV and $P_z \rightarrow \infty$ limits, is used 
to match these quasi parton distributions to ordinary parton distributions on the light-front~\cite{Ji:2013dva,Xiong:2013bka,Ma:2014jla}. 

This approach can be adopted to many observables, but one needs to first verify that the operator used in the
lattice calculation, when boosted to the infinite momentum frame (IMF), coincides with the one on the light-front.
The gluon helicity distribution is defined by Manohar in terms of the light-front correlation function~\cite{Manohar:1990kr}
\begin{equation}   \label{manohar}
\begin{aligned}
  \Delta G (x) S^+ &= \frac{i}{2xP^+} \int   \frac{\ud z^-}{2\pi} \,e^{ixP^+z^-}  \\
 &\hspace{-.5cm}\langle P,S| F^{+\alpha,a}(0) {\mathcal W}^{ab}(0,z^-)\tilde F^{+\phantom{\alpha},b}_{\phantom{+}\alpha} (z^-)|P,S\rangle,
\end{aligned}
\end{equation}
where ${\mathcal W}(0,z^-) = \mathcal P[e^{-ig\int^{z^-}_0\ud \zeta^- \,A^+(\zeta^-,\uvec 0_\perp)}]$ is a light-like Wilson line. After integrating over the longitudinal momentum fraction $x$, this light-front correlator reduces to the proper definition of gluon light-front helicity~\cite{Hatta:2011zs,Ji:2013fga}.
%
%\begin{equation} \label{eq4}
%S^G = \left[ \uvec{E}^a(0)\times [\uvec{A}^a(0)-\frac{1}{\nabla^+} 
%({\mathcal W}^{ab}\,\uvec{\nabla}A^{+,b})]\right]_z
%\end{equation}
Furthermore, it has recently been shown by Ji, Zhang and Zhao~\cite{Ji:2013fga} that this quantity coincides with the gluon spin contribution defined by Chen \emph{et al.}~\cite{Chen:2008ag,Chen:2009mr} in the IMF with $\uvec A_\text{phys}$ satisfying the non-Abelian Coulomb condition~\cite{Chen:2009mr}
\begin{equation}   \label{nacoul}
\uvec{\mathcal D}\cdot \uvec A_\text{phys}=0.
\end{equation}
 In other words, the longitudinal gluon spin operator turns into the helicity operator in the IMF. Similarly, Zhao, Liu and Yang~\cite{Zhao:2015kca} have shown that the quark and gluon canonical OAM defined by Chen \emph{et al.}~\cite{Chen:2008ag,Chen:2009mr} with the condition~\eqref{nacoul} in IMF coincide with the expressions derived explicitly by Hatta~\cite{Hatta:2011ku} and Lorc\'e~\cite{Lorce:2012ce} with staple light-like Wilson lines, namely
 \begin{align}
\lim_{P^z\to\infty} \langle L_{\text{can}}^{q,\,G}\rangle\Big|_{\uvec{\mathcal D}\cdot \uvec A_\text{phys}=0} &= \langle L^{q,G}_{\mathcal W_{\sqsupset}}\rangle.
\label{qgoam}
\end{align}

To carry out a lattice calculation of the canonical OAM and gluon spin, one possibility is to define 
$A^\mu_{\text{pure}}$ and
$A^\mu_{\text{phys}}$ with the constraint~\eqref{nacoul} on the lattice. It is shown in Refs.~\cite{Zhao:2015kca,Yang15} that their solutions are related to 
the gauge link $U^\mu(r)$ which connects the site $r$ to $r+a\hat \mu$ ($a$ is the lattice spacing). After the gauge 
link is fixed to that of the Coulomb gauge on the lattice with a gauge rotation matrix $g_c$
 \begin{equation}
 U^\mu(r)=g_c(r) U_c^{\mu}(r)g_c^{-1}(r+a\hat{\mu}),
\end{equation}
one can define the links $U^{\mu}_{\text{pure}}$ and $U_{\text{phys}}^\mu$ as
\begin{equation}
\begin{aligned}
U^{\mu}_{\text{pure}}(r)&\equiv g_c(r) g^{-1}_c(r+a\hat{\mu}),\\
U^{\mu}_{\text{phys}}(r)&\equiv U^\mu(r)-U^\mu_{\text{pure}}(r),
\end{aligned}
\end{equation}
so that
\begin{equation}
\begin{aligned}
A^{\mu}_{\text{pure}}(r)&=\tfrac{i}{ag}\,[U^{\mu}_{pure}(r)-\mathds{1}],\\
A^{\mu}_{\text{phys}}(r)&=g_c (r) A^\mu_c(r) g_c^{-1}(r) + O(a). 
\end{aligned}
\end{equation}
It is straightforward to prove~\cite{Yang15,Zhao:2015kca,Lorce:2012rr} that not only $A^\mu_{\text{pure}}$ and
$A^\mu_{\text{phys}}$ so constructed satisfy the gauge transformation properties in Eqs.~\eqref{inhomo} 
and~\eqref{homo}, but the pure gauge condition in Eq.~(\ref{puregaugecond})
and the non-Abelian Coulomb condition in Eq.~(\ref{nacoul}) are also satisfied to $\mathcal{O}(a)$. Note that $A^\mu_{\text{pure}}$ and $A^\mu_{\text{phys}}$ are solutions which are
obtained from the gauge rotation matrix $g_c(r)$ which fixes the Coulomb gauge and the Coulomb gauge-fixed
potential $A_c(r)$ itself. Therefore, one can construct the quark and gluon canonical OAM $\uvec L^q_\text{can}$ and $\uvec L^G_\text{can}$ as local operators on the lattice from $U^{\mu}_{pure}(r)$ and 
$A^\mu_{\text{phys}}(r)$. This strategy has recently been
applied to calculate the gluon light-front helicity $\Delta G$~\cite{Sufian:2014jma,Liu:2015nva}. 
\newline

Note that $\uvec L_{\text{can}}^q$ resembles the kinetic OAM in Eq.~\eqref{dec5}, except that the covariant derivative $D^\mu$ in the latter is replaced by $D^\mu_\text{pure}$. By analogy with Eq.~\eqref{q_contrib_def_1}, one might think to get 
$\uvec L^q_\text{can}$ by subtracting the quark spin from the quark total AM obtained from the symmetric canonical EMT
\begin{equation}
{\mathcal T}^{\mu\nu}_{\text{can},q}=\tfrac{i}{4}\sum_f \overline {\psi}_f 
    [\gamma^{\mu}\! \!\stackrel{\leftrightarrow}{D}\!\!\!\!\!\!\!\phantom{D}^{\nu}_\pure
     +\gamma^{\nu}\! \!\stackrel{\leftrightarrow}{D}\!\!\!\!\!\!\!\phantom{D}^{\mu}_\pure]  \psi_f.
\end{equation}
However, unlike the kinetic AM case in Eq.~\eqref{ang_mom_density}, the density of total AM is not simply related to this symmetric canonical EMT
\begin{equation}
M^{\mu\nu\alpha}_{\text{can},q}(r)\neq {\mathcal T}^{\mu\alpha}_{\text{can},q}r^\nu-{\mathcal T}^{\mu\nu}_{\text{can},q}r^\alpha,
\end{equation}
but involves extra terms.
Therefore, there is no such a relation as Eq.~\eqref{ang_op_def_split_3} in the canonical case. If, however, one adopts the asymmetric canonical EMT,

\begin{equation}
\widetilde{T}^{\mu\nu}_{\text{can},q}= \tfrac{i}{2}\sum_f \overline {\psi}_f \gamma^{\mu}\! \!\stackrel{\leftrightarrow}{D}\!\!\!\!\!\!\!\phantom{D}^{\nu}_\pure
      \psi_f,
\label{asymmetric}
\end{equation}
one does obtain, similarly to the kinetic case in Eq.~\eqref{kinOAMrel}, directly the quark canonical OAM
\begin{equation}\label{canOAMrel}
L^i_{\text{can},q}=\tfrac{1}{2}\,\epsilon^{ijk} \int \ud^3 r\, (\widetilde{T}^{0k}_{\text{can},q}r^j
-\widetilde{T}^{0j}_{\text{can},q}r^k).
\end{equation}
Because of the Stueckelberg/background dependence of the decomposition of the gauge potential~\eqref{Adec}, more structures in the parametrization of the off-forward matrix elements of  $\widetilde T^{\mu\nu}_{\text{can},q}$ are allowed. In particular, for the natural choice determined by the QCD factorization theorems, one has to introduce a light-like vector $n^\mu$. Once generalized
form factors are separated in lattice QCD calculations through proper linear combinations of matrix elements at different kinematics, then the quark canonical OAM can be obtained through the relation~\cite{Lorce:2015lna}
\begin{equation}
 \langle L^{q}_\text{can}\rangle=A^{e,q}_4(0,0).
\end{equation}
Other similar relations have also been obtained for the gluon canonical OAM and the quark/gluon spin contributions.

Alternatively, one can directly calculate the canonical OAM from
the off-forward matrix element, just like the magnetic moment can be obtained from the off-forward spin-flip matrix element~\cite{Alexandrou:2014exa},
\begin{equation}
\begin{aligned}
&\epsilon^{ijk} \langle p,S|\int \ud^3r\,\overline{\psi}\gamma^0 r^j iD_\text{pure}^k \psi |p,S\rangle=\\
&\epsilon^{ijk}\lim_{\Delta \rightarrow 0} \langle p',S' |\int \ud^3r\, \overline{\psi}\gamma^0 \tfrac{\partial}{\partial \Delta^j}
e^{i \uvec{\Delta}\cdot\uvec{r}} D^k_\text{pure}\psi |P,S\rangle.
\end{aligned}
 \end{equation}
The challenge here is to have a reliable extrapolation to the $\Delta \to 0$ limit which can be ameliorated with
a large lattice or twisted boundary condition to have data at small $\Delta$.
\newline

The third option is the calculation of quark and gluon OAM from GTMDs. Exploratory lattice calculation of TMDs have been carried out in Refs.~\cite{Musch:2010ka,Musch:2011er,Engelhardt:2014eea,Engelhardt:2015xja} where a staple-like Wilson line $\mathcal W_{\sqsupset}$ was used in the calculation and the result were evolved to the IMF to obtain the light-front correlator.  This approach can
be generalized to GTMDs and use Eq.~\eqref{LzGTMD} to obtain the quark and gluon canonical OAM. Since
the link between the bilinear quark fields is non-local, one of the challenges in this approach is the 
renormalization of the non-local operator on the lattice.

\section{Effective theory of baryons}\label{sec6}

Many estimates of quark spin and OAM contributions of the nucleon are based on quark models. However, quark models are not realistic effective theories of QCD, since they do not have chiral symmetry, a salient feature of QCD whose dynamics governs light-quark hadron structure, spectroscopy, and scattering at low energies. It is being learned quantitatively through lattice calculations of quark spin from the anomalous Ward 
identity~\cite{Yang:2015xga,Liu:2015nva} that the smallness of the quark spin contribution in the nucleon is mainly due to the $U(1)$ anomaly, the same origin which is responsible for the large $\eta'$ mass. 
This cannot be understood with quark models without the chiral $U(1)$ anomaly. Similarly, relativistic quark models do not explain the large OAM obtained from the lattice calculation in Sec.~\ref{lattice_results}. Both the chiral quark model studies~\cite{Glozman:1995fu} and lattice
calculation of valence QCD~\cite{Liu:1998um,Liu:2014jua} reveal that the level reversal of the positive and negative parity excited states of the nucleon, \emph{i.e.} $P_{11}(1440)$ (Roper resonance) and $S_{11}(1535)$, and the hyperfine splittings between the decuplet and octet baryons are dominated by the meson-mediated flavor-spin interaction, not the gluon-mediated color-spin interaction. 
All of these point to the importance of the meson degree of freedom  ($q\overline{q}$ pairs in the higher Fock space) which is missing in the quark model. 

To see how this comes about, one can follow Wilson's renormalization group approach to effective theories~\cite{Wilson:1973jj,Polchinski:1983gv}. It is 
suggested by Liu {\it et al.}~\cite{Liu:1999kq} that the effective theory for baryons between the scale of 
$4\pi f_{\pi}$, which is the scale of the meson size ($l_M\sim 0.2$ fm), and $\sim 300$ MeV, which is the
scale of a baryon size ($l_B\sim 0.6$ fm), should be a chiral quark model with renormalized couplings 
and renormalized meson, quark and gluon fields which preserve chiral symmetry. A schematic illustration
for such division of scales\footnote{We should point out that although
two scales are adopted here, they are distinct from those of
Manohar and Georgi~\cite{Manohar:1983md}. In the latter, the $\sigma$ -- quark model
does not make a distinction between the quark fields in the baryons and mesons. As such,
there is an ambiguity of double counting of mesons and
$q\overline{q}$ states. By making the quark-quark confinement length scale
$l_B$ larger than the quark-antiquark confinement length scale $l_M$, one
does not have this ambiguity.} for QCD effective theories is illustrated in Fig.~\ref{2scales}.

\begin{figure}[t]
\begin{center}
\includegraphics[width=0.3\textwidth]{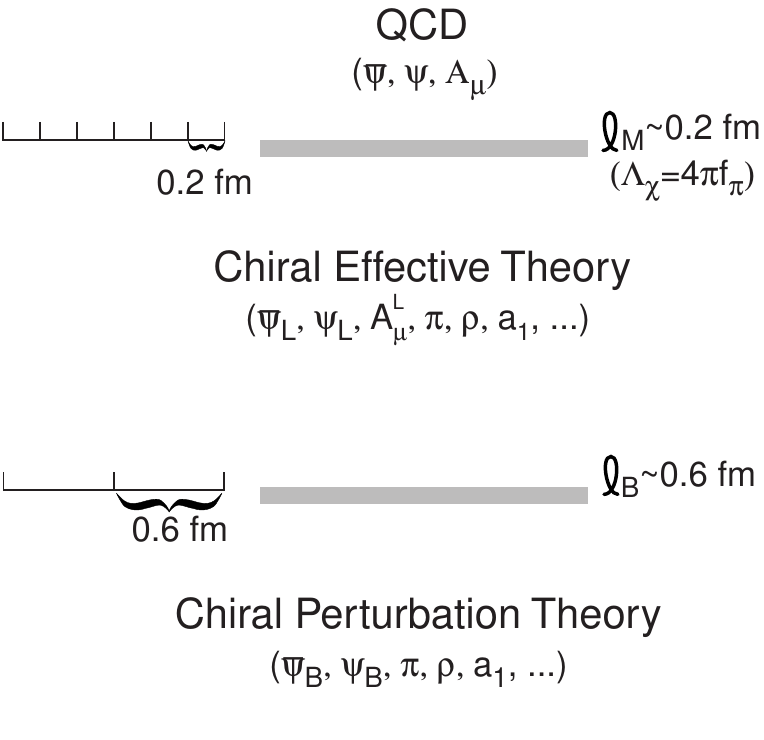}
\end{center}
\caption{A schematic illustration of the the two-scale delineation of
the effective theories. The shaded bars mark the positions of the 
cutoff scales $l_M$ and $l_B$  separating
different effective theories.}
\label{2scales}
\end{figure}

This suggestion is based on the observation that mesons and baryon form factor assume a monopole and dipole form, respectively. Since
the $\pi NN$ form factor is much softer than the $\rho \pi \pi$
form factor, it is suggested that the confinement scale of quarks in the
baryon $l_B$ is larger than $l_M$ -- the confinement scale between the 
quark and antiquark in the meson.
This is consistent with the large-$N_c$ approach to hadrons where the mesons 
are treated as point-like fields and the baryons emerge as solitons
with a size of order unity in $N_c$~\cite{Witten:1979kh}.
Taking $l_M$ from the $\rho \pi \pi$ form factor gives
$l_M \sim 0.2$ fm. This is very close to the chiral symmetry
breaking scale set by $\Lambda_{\chi} = 4 \pi f_{\pi}$. Considering
them to be the same, operators of 
low-lying meson fields become relevant operators below $\Lambda_{\chi}$. As for the baryon confinement
scale, Liu {\it et al.} take it to be the size characterizing the meson-baryon-baryon
form factors. Defining the latters from the respective meson poles in the nucleon pseudoscalar, 
vector, and axial-vector form factors in a lattice calculation~\cite{Liu:1998um} (see Fig. 17 
in the reference), Liu {\it et al.} obtained $l_B \sim 0.6 - 0.7$ fm, satisfying $l_B> l_M$. Thus, in between these two scales $l_M$ and
$l_B$, one could have coexistence of mesons and quarks in an effective theory for baryons.

An outline is given by Liu {\it et al.}~\cite{Liu:1999kq} to show how to construct a chiral 
effective theory for baryons. In the intermediate length scale between $l_M$ and $l_B$,
one needs to separate the fermion and gauge fields into long-range ones and
short-range ones
\begin{equation}
\psi = \psi_L + \psi_S, 
\qquad  A^{\mu} = A_L^{\mu} + A_S^{\mu},
\end{equation}
where $\psi_L/\psi_S$ and $A_L^{\mu}/A_S^{\mu}$ represent the infrared/ultraviolet 
part of the quark and gauge fields, respectively, with momentum 
components below/above $1/l_M$ or $\Lambda_{\chi}$. One adds irrelevant higher-dimensional operators to the
ordinary QCD Lagrangian 
with coupling between bilinear quark fields and auxiliary fields as
given in Ref.~\cite{Li:1995aw}, interpreting these quark fields 
as the short-range ones, \emph{i.e.} $\psi_S$ and $\overline{\psi}_S$.
Following the procedure by Li in Ref.~\cite{Li:1995aw}, one can integrate out the
 short-range fields and perform the derivative expansion to
bosonize $\psi_S$ and $\overline \psi_S$. This leads to
the Lagrangian with the following generic form:
\begin{equation}  \label{ect}
\begin{aligned}
{\cal L}_{\chi QCD} &= {\cal L}_{QCD'}(\overline{\psi}_L, \psi_L, A^{\mu}_L)
+ {\cal L}_M(\pi, \sigma, \rho, a_1, G,\cdots)  \\
&\quad+ {\cal L}_{\sigma q}
(\overline{\psi}_L, \psi_L, \pi, \sigma, \rho, a_1, G, \cdots). 
\end{aligned}
\end{equation}
${\cal L}_{QCD'}$ includes the original form of QCD but in terms of 
the quark fields $\overline{\psi}_L, \psi_L$, and
the long-range gauge field $ A^{\mu}_L$ with renormalized couplings. It 
also includes higher-order covariant derivatives~\cite{Warr:1986we}. 
${\cal L}_M$ is the meson effective
Lagrangian, \emph{e.g.} the one derived by Li~\cite{Li:1995aw} which should include the
glueball field $G$. Finally, ${\cal L}_{\sigma q}$ gives the coupling
between the $\overline{\psi}_L$, $\psi_L$, $G$ and mesons. As we see, in this
intermediate scale, the quarks, gluons and mesons coexist and
meson fields couple to the long-range quark fields.
Going further down below the baryon confinement scale $1/l_B$, one
can integrate out $\overline{\psi}_L$, $\psi_L$ and $A^{\mu}_L$, resulting in
an effective Lagrangian ${\cal L}(\overline{\Psi}_B, \Psi_B, \pi, \sigma, \rho, 
a_1, G,\cdots)$ in terms of the baryon and meson fields~\cite{Wang:1999xh}. This would 
correspond to an effective theory in the chiral perturbation theory.
In order for the chiral symmetry to be preserved, the effective theory of baryons at
the intermediate scale necessarily involves mesons in addition to the effective
quark and gluon fields. This naturally leads to a chiral
quark effective theory. 

Models like the little bag model with skyrmion outside the MIT bag~\cite{Brown:1979ui}, the cloudy bag 
model~\cite{Thomas:1981vc} and quark chiral soliton model~\cite{Wakamatsu:2006dy} have
the right degrees of freedom and, thus, could 
possibly delineate the pattern of division of the proton spin with large 
quark OAM contribution. In particular, the fact that the $u$ and $d$ OAM in the MIT bag and to some extent the LFCQM in Table 1 have different signs from those of the lattice calculation may well be due to the lack of meson contributions as demanded by chiral symmetry in the effective theory of baryons.

\section{Spin-orbit correlation}\label{sec7}

We have seen that parton OAM constitutes an important part of the proton spin puzzle. Lorc\'e and Pasquini~\cite{Lorce:2011kd} showed that another interesting and independent information about the nucleon spin structure is encoded in the correlation between the parton OAM and its own polarization. This quantity escaped attention because it does not contribute to the nucleon spin budget, but naturally appears once one considers all correlations allowed by parity. Parton spin-orbit correlation is in some sense the parity partner of parton OAM~\cite{Lorce:2014mxa}. It is therefore hardly surprising that most of the results and discussions we had about parton OAM find a counterpart for parton spin-orbit correlation.

For example, the quark kinetic and canonical spin-orbit operators are defined as~\cite{Lorce:2011kd,Lorce:2014mxa}
\begin{equation}
\begin{aligned}
\uvec C^q_\text{kin}&=\int\ud^3r\,\psi^\dag\gamma_5(\uvec r\times i\uvec D)\psi,\\
\uvec C^q_\text{can}&=\int\ud^3r\,\psi^\dag\gamma_5(\uvec r\times i\uvec D_\pure)\psi.
\end{aligned}
\end{equation}
Like the average quark kinetic OAM, the average quark kinetic spin-orbit correlation can be expressed in terms of moments of twist-2 and twist-3 GPDs~\cite{Lorce:2014mxa}
\begin{align}\label{SOtwist2}
\langle C^q_\text{kin}\rangle&=\tfrac{1}{2}\int\ud x\,x\tilde H^q(x,0,0)-\tfrac{1}{2}\,[F^q_1(0)-\tfrac{m_q}{2M_N}\,H^q_1(0)],\nonumber\\
&=-\int\ud x\,x[\tilde G^q_2(x,0,0)+2\tilde G^q_4(x,0,0)].
\end{align}
The parton spin-orbit correlation can also be naturally expressed as a phase-space integral~\cite{Lorce:2011kd,Lorce:2014mxa}
\begin{equation}\label{SOWigner}
\begin{aligned}
&\langle C^{q,G}_{\mathcal W}\rangle=\\
&\quad\int\ud x\,\ud^2k_\perp\,\ud^2b_\perp\,(\vec b_\perp\times\vec k_\perp)_z\,\tilde\rho^{q,G}_{++}(x,\vec k_\perp,\vec b_\perp;\mathcal W),
\end{aligned}
\end{equation}
where $\tilde\rho^{q,G}_{++}(x,\vec k_\perp,\vec b_\perp;\mathcal W)$ is the helicity phase-space distribution describing the difference between the distributions of quarks/gluons with polarization parallel and antiparallel to the longitudinal direction. In terms of the GTMDs, this relation reads~\cite{Lorce:2011kd,Kanazawa:2014nha,Lorce:2014mxa}
\begin{equation}
\langle C^{q,G}_{\mathcal W}\rangle=\int\ud x\,\ud^2k_\perp\,\tfrac{\uvec k^2_\perp}{M^2}\,G^{q,G}_{11}(x,0,\uvec k_\perp,\uvec 0_\perp;\mathcal W),
\end{equation}
where once again the shape of the Wilson line $\mathcal W$ determines whether the spin-orbit correlation is of the kinetic or canonical type.

Like in the OAM case, the integrand of Eq.~\eqref{SOWigner} itself can consistently be interpreted as the phase-space density of quark/gluon spin-orbit correlation
\begin{equation}\label{SOdensity}
\langle C^{q,G}_{\mathcal W}\rangle(x,\uvec k_\perp,\uvec b_\perp)=(\uvec b_\perp\times\uvec k_\perp)_z\,\tilde\rho^{q,G}_{++}(x,\uvec k_\perp,\uvec b_\perp;\mathcal W).
\end{equation}
From parity and time-reversal symmetries, it follows that the helicity phase-space distribution of longitudinally polarized quarks/gluons inside a longitudinally polarized nucleon can be decomposed into four contributions
\begin{equation}
\begin{aligned}
\tilde\rho^{q,G}_{++}&=\tilde\rho^{q,G}_1+(\uvec b_\perp\cdot\uvec k_\perp)\,\tilde\rho^{q,G}_2+(\uvec b_\perp\times\uvec k_\perp)_z\,\tilde\rho^{q,G}_3\\
&\quad+(\uvec b_\perp\times\uvec k_\perp)_z\,(\uvec b_\perp\cdot\uvec k_\perp)\,\tilde\rho^{q,G}_4,
\end{aligned}
\end{equation}
where $\tilde\rho^{q,G}_i\equiv\tilde\rho^{q,G}_i(x,\uvec k^2_\perp,(\uvec b_\perp\cdot\uvec k_\perp)^2,\uvec b^2_\perp;\mathcal W)$. The functions $\tilde\rho_1$ and $\tilde\rho_2$ are related to the Fourier transform of the GTMD $G_{14}$, whereas $\tilde\rho_3$ and $\tilde\rho_4$ are related to the Fourier transform of the GTMD $G_{11}$. Once again, the factor $(\uvec b_\perp\cdot\uvec k_\perp)$ indicates that $\tilde\rho_2$ and $\tilde\rho_4$ are naive $\mathsf T$-odd, \emph{i.e.} they change sign under $\mathcal W_{\sqsupset}\leftrightarrow\mathcal W_{\sqsubset}$, and arise therefore from the imaginary part of the corresponding GTMDs. Note also that the only contribution to the spin-orbit correlation surviving integration over $\uvec b_\perp$ or $\uvec k_\perp$ is $\tilde\rho_3$. For a straight Wilson line $\mathcal W_\mid$, one obtains the density of JXY quark and gluon OAM~\cite{Ji:2012sj,Ji:2012ba} provided that one integrates at least over $\uvec k_\perp$~\cite{Lorce:2012ce}.

Using the extraction of polarized quark distributions from combined inclusive and semi-inclusive DIS data~\cite{Leader:2010rb} in very good agreement with lattice calculations~\cite{Bratt:2010jn}, Lorc\'e found the estimates $\langle C^u_\text{kin}\rangle\approx -0.8$ and $\langle C^d_\text{kin}\rangle\approx -0.55$, meaning that quark spin and kinetic OAM are in average antiparallel. On the contrary, the canonical version of the quark spin-orbit correlation appears to be positive in light-font quark models~\cite{Lorce:2011kd}, illustrating the importance of the quark-gluon interaction. Here also it would be very interesting to compute these quantities using the large momentum approach of lattice gauge theory.

\section{Summary}\label{sec8}

%### Comment: this is a partial conclusion. Need to add more material.
We reviewed the recent development on the kinetic and canonical decompositions of the proton spin addressing
the issues of gauge invariance and, in particular, their relevance to the parton description of the orbital angular momentum (OAM) from generalized parton dsitributions (GPDs), transverse-momentum dependent parton distributions (TMDs), and  generalized TMDs (GTMDs).  Some of the theoretical questions and puzzles are also discussed.

In order to have concrete results on OAM and other contributions to the proton spin, it is imperative to carry out lattice calculations with control of both statistical and systematic errors in order to compare with experiments when they are available from GPDs or TMDs and gain insight on the nature of the decomposition of the proton spin into its quark and gluon components. To this aim, the status and challenges of lattice calculations on the OAM were summarized. 

Finally, we pointed out that there is an additional information about the proton spin structure which is encoded in the 
spin-orbit correlation and which deserves theoretical and experimental explorations.

\begin{acknowledgement}
This work is is supported partially by US Department of Energy grant DE-SC0013065.
C.L. acknowledges support from the Belgian Fund F.R.S.-FNRS \emph{via} the contract of Charg\'e de Recherches.
\end{acknowledgement}

%\begin{figure}
%\resizebox{0.75\textwidth}{!}{%
%  \includegraphics{leer.eps}
%\vspace{5cm}       % Give the correct figure height in cm
%\end{figure}
%\begin{figure*}
%\vspace*{5cm}       % Give the correct figure height in cm
%\end{figure*}
%\begin{table}
%\begin{tabular}{lll}
%\hline\noalign{\smallskip}
%\end{tabular}
% Or use
%\vspace*{5cm}  % with the correct table height
%\end{table}

\end{document}